\documentclass[journal ]{IEEEtran}
\usepackage{lineno}
\usepackage{amsmath,amssymb}
\usepackage{float}

\usepackage{makecell}
\usepackage{CJK} 
\usepackage{textcomp}
\usepackage{float}
\usepackage{multirow}
\usepackage{graphicx}
\usepackage{multirow}
\usepackage{tabu,bm}
\usepackage{color}
\usepackage{subfigure}
\usepackage{booktabs}

\usepackage[linesnumbered,ruled,vlined]{algorithm2e}
\usepackage{algorithmic}
\usepackage{url}

\usepackage{cite}



\begin{document}
%
\title{Task-Oriented Multi-User Semantic Communications}
%
%
%

\author{Huiqiang Xie,~\IEEEmembership{Student Member,~IEEE,}
Zhijin~Qin, \IEEEmembership{Senior~Member,~IEEE,}
Xiaoming~Tao, \IEEEmembership{Member,~IEEE,}\\
and Khaled~B.~Letaief, \IEEEmembership{Fellow,~IEEE}
\thanks{Huiqiang Xie and Zhijin Qin are with the School of Electronic Engineering and Computer Science, Queen Mary University of London, London E1 4NS, UK (e-mail: h.xie@qmul.ac.uk, z.qin@qmul.ac.uk).}
\thanks{Xiaoming Tao is with  the Department of Electronic Engineering and also with the Beijing National Research Center for Information Science and Technology, Tsinghua University, Beijing 100084, China (e-mail: taoxm@tsinghua.edu.cn).}
\thanks{Khaled B. Letaief is with the Department of Electronic and Computer
Engineering, The Hong Kong University of Science and Technology, Hong
Kong, and also with Peng Cheng Laboratory, Shenzhen 518066, China (e-
mail: eekhaled@ust.hk).}
}
\maketitle

\begin{abstract}
While semantic communications have shown the potential in the case of single-modal single-users, its applications to the multi-user scenario remain limited. In this paper, we investigate deep learning (DL) based  multi-user semantic communication systems for transmitting single-modal data and multimodal data, respectively. We will adopt three intelligent tasks, including, image retrieval, machine translation, and visual question answering (VQA) as the transmission goal of semantic communication systems. We will then propose a Transformer based unique framework to unify the structure of transmitters for different tasks. For the single-modal multi-user system, we will propose two Transformer based models, named, DeepSC-IR and DeepSC-MT, to perform image retrieval and machine translation, respectively.  In this case, DeepSC-IR is trained to optimize the distance in embedding space between images and DeepSC-MT is trained to minimize the semantic errors by recovering the semantic meaning of sentences. For the multimodal multi-user system, we develop a Transformer enabled model, named, DeepSC-VQA, for the VQA task by extracting text-image information at the transmitters and fusing it at the receiver. In particular, a novel layer-wise Transformer is designed to help fuse multimodal data by adding connection between each of the encoder and decoder layers. Numerical results will show that the proposed models are superior to traditional communications in terms of the robustness to channels, computational complexity, transmission delay, and the task-execution performance at various task-specific metrics.
\end{abstract}


%
\IEEEpeerreviewmaketitle

\section{Introduction}
%
%
%
%


Conventional communication systems are regarded as transmission pipes, in which the data are collected at the transmitters and reconstructed at the receivers. As we step into the era of connected intelligence \cite{LetaiefCSZZ19}, the widely deployed devices have been generating unprecedented amounts of multimodal data to serve various tasks, which makes conventional communications a new bottleneck and performance limit. There exist two ways to address this problem: 1) evolution of hardware to enlarge the system capacity and transmission rate, e.g., millimeter wave/terahertz communications~\cite{wang2015multi, han2018propagation}, massive antennas array~\cite{LuLSAZ14}, and reconfigurable intelligent surfaces~\cite{HuangZADY19}; 2) improvement of software to optimize the utilization of communication resources, e.g., data compression~\cite{sayood2017introduction}, resource multiplexing~\cite{goldsmith2005wireless}, and semantic communications~\cite{KountourisP21}. In this work, we investigate the second approach. Moreover, we will focus on semantic communications,  the new emerging communication paradigm, which has shown its superiority in handling the massive volume of data.  

Semantic communications are content-aware, task-oriented, and semantic-related, in which only important, relevant, and useful information to the users/applications are extracted from a large amount of data and delivered to the destinations. The existing works on semantic communications can be categorized into two parts: full data reconstruction and  task execution.

For the data reconstruction, semantic communications generally extract the global semantic information behind  data and reconstruct the data based on the received semantic information. Farsad~\textit{et al.}~\cite{FarsadRG18} designed the initial deep joint source-channel coding for text transmission, in which the text sentences are encoded into fixed-length bit streams over simple channel environments. With the depth exploration in the semantic communications, Xie~\textit{et al.}~\cite{TSP-XieQLJ21} developed more powerful joint semantic-channel coding, named DeepSC, to encode text information into various length over complex channels. Moreover, Xie~\textit{et al.}~\cite{JSAC-XieQ21} also proposed an environment-friendly semantic communication system, named L-DeepSC, for the capacity-limited devices. Besides, Bourtsoulatze~\textit{et al.}~\cite{tccn-BourtsoulatzeKG19} investigated the initial deep image transmission semantic communication systems, in which the semantic and channel coding are optimized jointly. Kurka~\textit{et al.} \cite{jsait-KurkaG20} extended Bourtsoulatze's work with the channel feedback to improve the quality of image reconstruction. Weng~\textit{et al.} \cite{jsac-WengQ21} developed an attention mechanism based semantic communication systems to reconstruct speech signals.

For the task-specific applications, only the semantic information useful for serving the task execution is extracted at the transmitter, which will be directly used for the decision making at the receiver. Lee \textit{et al.} \cite{access-LeeLCC19} developed an image classification-oriented semantic communications for improving the recognition accuracy rather than performing image reconstruction and classification separately. Jankowski \textit{et al.}~\cite{jsac-JankowskiGM21} considered image based re-identification for person or cars as the communication task, in which two schemes (digital and analog) are proposed to improve the retrieval accuracy. Except from image based tasks, Weng \textit{et al.}~\cite{weng2021semantic} designed speech recognition-oriented semantic communications, named, DeepSC-SR, to directly recognize the speech signals into texts.  The prior works explore the possibility of semantic communications for transmitting signals in a single-modal single-user system. However, in practice, we must gather multimodal data from different users/devices, transmit over the air, and process/fuse multimodal data at the receiver. This motivates us to develop a multi-user semantic communication system to support multimodal data transmission. Our initial design of the MU-DeepSC is for serving the visual question answering (VQA) task to improve the answer accuracy \cite{xie2021task}, which adopts Long Short Term Memory (LSTM) for the text transmitter and Convolutional Neural Network (CNN) for the image transmitter. However, a unified framework to support various tasks with multimodal data is still missing in multi-user semantic communications.

Particularly, single-modal multi-user semantic communications represent the extension of single-modal single-user semantic communications, in which multiple single-modal intelligent tasks can be performed simultaneously but each user is only associated with one intelligent task. Multimodal multi-user semantic communications employ more than one user to serve one multimodal intelligent task, which is suitable for the emerging autonomous scenarios in daily life \cite{ramachandram2017deep} and industry \cite{vakil2021survey}, i.e., autonomous checkout at retail stores~\cite{ruiz2019autotag}, intelligent control at smart home~\cite{stojkoska2017review}, and human activity recognition in smart healthcare~\cite{cvpr-ZouYDLZS19}. Such scenarios are achieved by collecting multimodal data from the various sensors so as to provide the information in a complementary manner and fuse them at the server/cloud.  For the design of multi-user semantic communications, we face the following challenges:
\begin{enumerate}
    \item[\textit{Q1}:] \textit{How to extract semantic information at the transmitter for both single-modal and multimodal multi-user semantic communications?}
    \item[\textit{Q2}:] \textit{How to reduce the interference from other users for both single-modal and multimodal multi-user semantic communications?}
    \item[\textit{Q3}:] \textit{How to process/fuse the received semantic information at the receiver for  multi-user semantic communications to transmit multimodal data?} 
\end{enumerate}

In this paper, we investigate task-oriented multi-user semantic communications for transmitting data with single modality and multiple modalities by considering two types of sources: image and text. We choose image retrieval and machine translation for transmission data with single-modality, and one of the most challenging tasks, namely, the visual question answering (VQA) task, for illustrating transmission with multimodal data. The main contributions of this paper are summarized as follows:
\begin{itemize}
    \item We propose a Transformer \cite{vaswani2017attention} based transmitter structure, which is applicable for both text and image transmission by effectively extracting semantic information for different tasks. This addresses the aforementioned \textit{Q1}.
    \item We demonstrate the efficient methods for training the proposed structure. In particular, the transmitters and receiver in the proposed frameworks are trained jointly to eliminate distortion from the channels and interference from other users. This addresses the aforementioned \textit{Q2}.
    \item Based on the proposed structure, we propose three different deep learning (DL) enabled multiuser semantic communication frameworks, named DeepSC-IR for image retrieval, DeepSC-MT for machine translation, and DeepSC-VQA for VQA. Specially, we propose a novel layer-wise Transformer, which can exploit more text information to guide image information, to fuse the text and image information. This addresses the aforementioned \textit{Q3}.
    \item  Based on extensive simulation results, the proposed frameworks outperform the traditional communication systems with lower requirements on the communication resources and improved system robustness at the low SNR regimes.
\end{itemize}

The rest of this paper is organized as follows. The related works of selected tasks and preliminaries are briefly reviewed in Section II. The system model is introduced in Section III.  The proposed single-modal multi-user semantic communications are proposed in Section IV. Section V details the proposed multimodal multi-user semantic communications. Numerical results are presented in Section VI to show the performance of the proposed frameworks. Finally, Section VII concludes this paper.

\textit{Notation}: $\mathbb{C}^{n \times m}$ and $\mathbb{R}^{n \times m}$ represent sets of complex and real matrices of size  $n \times m$, respectively. Bold-font variables denote matrices or vectors. $x \sim {\cal CN}(\mu,\sigma^2)$ means variable $x$ follows a circularly-symmetric complex Gaussian distribution with mean $\mu$ and covariance $\sigma^2$. $(\cdot)^{\text T}$ and $(\cdot)^{\text H}$ denote the transpose and Hermitian, respectively. $\Re \{\cdot \}$ and $\Im \{\cdot \}$ refer to the real and imaginary parts of a complex number.  

\section{Related Works and Preliminaries}
In this section, we will first introduce the definitions of the three intelligent tasks, including image retrieval, machine translation, and VQA. We then briefly review the related works on the three tasks. Because the designed models in the next sections mainly consist of the Transformer network, we will briefly introduce the preprocessing for image and text, and the main components for the Transformer network.

\subsection{Image Retrieval} 
The image retrieval task aims to identify the top-$k$ similar images by matching the sent image with those stored in a large server, and returns the similar ones to users. For example, the user uploads a dress image to  Amazon app and wishes to find similar dress products. Such image retrieval tasks cannot be performed locally due to the centralized database. 

Modern methods for image retrieval typically rely on DL based models by extracting compact image-level features~\cite{chen2021deep} for image match or classification.  Recent techniques mainly focus on two parts: deep network architectures and training algorithms. The deep network architectures include single feedforward pass models \cite{tolias2016particular}, multiple feedforward pass models \cite{gordo2016deep}, attention based models \cite{babenko2015aggregating}, and deep hashing embedding based models~\cite{HuangYPL18}. While the training algorithms focus on classification based learning~\cite{teh2020proxynca}, metric based learning~\cite{el2021training}, and unsupervised-based learning~\cite{gu2019clustering}.

\subsection{Machine Translation}
One core of communications is to transmit the meanings behind the text, however, one of the major obstructs for communications is the different grammars and presentations for different languages. Therefore, for the machine translation task, the intention is that the transmitter sends one language, i.e., Chinese, and the receiver directly receives the desired language, i.e., English, which aims to broken the obstruct of communications and improve communication efficiency. 

The recent successful approaches for machine translation problems are mostly based on the classic encoder-decoder structure \cite{kalchbrenner2013recurrent}, in which the encoder extracts the sentence-level intermediate features at source language and the decoder provides the entire sentence at target language based on the intermediate features. The representative models include CNN based models \cite{MengLWLJL15}, Transformer based models \cite{vaswani2017attention}, and RNN based models, i.e., LSTM networks \cite{hochreiter1997long} and Gated Recurrent Units (GRU) networks \cite{ChoMGBBSB14}.

\subsection{Visual Question Answering}
In the VQA task, the semantic information from different users is correlated. One user may transmit the vision information collected by a camera while the other user sends the text information collected by a sensor. Then, the transmitted vision and text information from different users is employed to carry out the answers at the receiver. 

The core of VQA tasks is multimodal data fusion techniques~\cite{zhou2015simple}, in which the image and questions in text are first represented as global features and then fused by a multimodal fusion model to predict the answer. Recent approaches adopt the visual attention mechanism by attending image features with given question features, which include multimodal bilinear pooling methods~\cite{KimOLKHZ17}, stacked attention network~\cite{YangHGDS16}, bottom-up and top-down attention mechanism~\cite{00010BT0GZ18}, and co-attention network~\cite{Yu0CT019}.

\subsection{Preliminaries}

The \textit{text preprocessing} includes two parts: tokenize and embedding. The input sentence is first splitted into scalar-wise tokens, each  representing one word or one sub-word. These scalar-wise tokens are then mapped into vector-shaped tokens with learnable word vectors and used as the input to the Transformer. The \textit{image preprocessing} also includes two parts: patchify and project. The input image is first decomposed into fixed-sized patches, e.g. 16x16. Each patch is linearly projected into vector-shaped tokens and used as an input to the Transformer. An extra learnable $<$CLS$>$ token is added to the input sequence such that its corresponding output token serves as a global representation for the input sequence. The location prior is incorporated by adding a learnable one-dimension (1-D) positional encoding vector to the input tokens. 

Transformer network consists of the encoder layers and decoder layers. Each encoder layer includes two main blocks: 1) a Multi-Headed Self Attention layer, which applies a self-attention operation to different projections of input tokens; and 2) a Feed-Forward layer. The decoder layer includes three main blocks: 1) a Multi-Headed Self Attention layer; 2) a Multi-Headed Guided Attention layer, which applies a attention operation to the projections of input tokens and the output tokens of encoder; and 3) a Feed-Forward layer. All blocks are preceded by layer normalization and followed by a skip connection.

\begin{figure*}
    \centering
    \includegraphics[width=150mm]{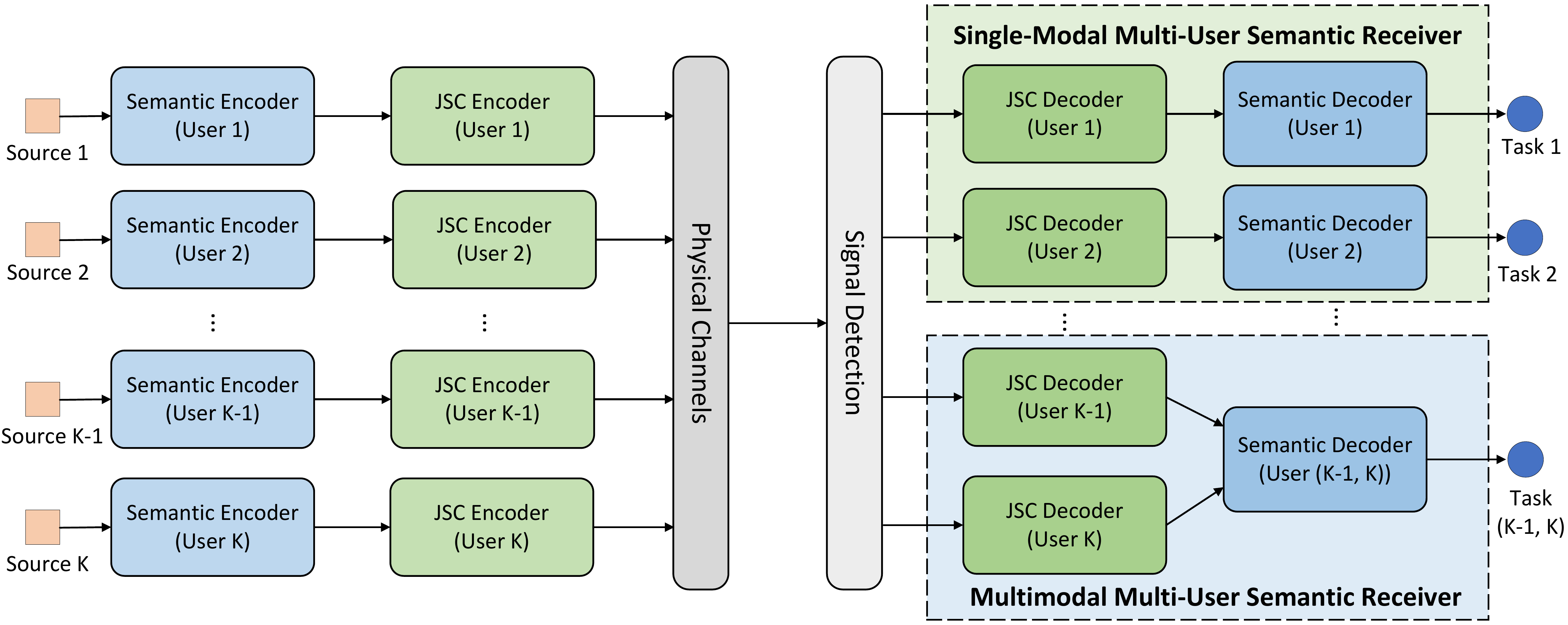}
    \caption{The framework of multi-user semantic communication systems}
    \label{fig:system-model}
\end{figure*}

\section{System Model}
As shown in Fig. \ref{fig:system-model}, we consider the multi-user semantic communication system, which consists of one receiver equipped with $M$ antennas and $K$ single-antenna transmitters. We will focus on the multi-user semantic communication system with single-modal data and multimodal data to transmit, respectively. The single-modal multi-user scenario means that each user transmits independent semantic information to perform its own task. The multimodal multi-user scenario indicates that the data from different users are semantically complementary. 

\subsection{Semantic Transmitter}
As shown in Fig. \ref{fig:system-model},  we denote the source data of the $k$-th user as ${\bm s}^{\cal Q}_k$ with modality ${\cal Q} \subseteq \left\{ { {\cal I}:image, {\cal T}:text, {\cal V}:video, {\cal S}:speech} \right\}$, where each source contains the semantic information. The semantic information is extracted first by 
\begin{equation}
    {\bm z}^{\cal Q}_k={S\left( {{{\bm s}^{\cal Q}_k};{{\bm{\alpha }}^{\cal Q}_k}} \right)},
\end{equation}
where ${\bm z}^{\cal Q}_k \in \mathbb{R}^{L_S \times 1}$ is the semantic information with length $L_S$ and ${S\left( {;{{\bm{\alpha }}^{\cal Q}_k}} \right)}$ is the modality $\cal Q$ semantic encoder for the $k$-th user with learnable parameters ${{\bm{\alpha }}^{\cal Q}_k}$. Due to the limited communication resource and complex communication environment for wireless communications, the semantic information of the $k$-th user is compressed by
\begin{equation}
    {{\bm x}^{\cal Q}_k} = C\left( {{\bm z}^{\cal Q}_k;{{\bm{\beta }}^{\cal Q}_k}} \right),
\end{equation}
where ${\bm x}^{\cal Q}_k \in {\mathbb{C} }^{L_C \times 1}$ is the transmitted complex signal with length $L_C < L_S$  and ${C\left( { ;{{\bm{\beta }}_k}} \right)}$ is the $k$-th user  {joint source-channel (JSC)} encoder for modality $\cal Q$ with learnable parameters, ${{\bm{\beta }}_k}$.  {The neural JSC encoder in semantic communications compresses semantic information to reduce the number of transmitted symbols, as well as improve the robustness to channel variations.}

\subsection{Semantic Receiver}
When the transmitted signal passes a multiple-input multiple-output (MIMO) physical channel, the received signal, ${\bf Y} \in {\mathbb C}^{M \times L_C}$, at the receiver can be expressed as
\begin{equation}\label{eq3}
    {\bf Y} = {\bf H \bf X} + {\bf N},
\end{equation}
where ${\bf X}^T=\left [{\bm x}^{\cal Q}_{1}, {\bm x}^{\cal Q}_{2},\cdots,{\bm x}^{\cal Q}_{K} \right ]\in {\mathbb{C} }^{ L_C \times K}$ denotes transmit symbols from all $K$ users,  ${\bf H}=\left [{\bm h}_{1},{\bm h}_{2},...,{\bm h}_{K} \right ]\in {\mathbb{C} }^{M\times K}$ is the channel matrix between the BS and users.  {For the Rayleigh fading channel, the channel coefficient follows ${\cal CN}(0,1)$; for the Rician fading channel, it follows ${\cal CN}(\mu,\sigma^2)$ with $\mu = \sqrt{r/(r+1)}$ and $\sigma = \sqrt{1/(r+1)}$, where $r$ is the Rician coefficient.}  $ {\bf N}\in {\mathbb{C} }^{M\times L_C}$ denotes the circular symmetric Gaussian noise. The elements of $\bf N$ are i.i.d with zero mean and variance ${\sigma }_{n}^{2}$, and SNR is defined by $\sum\limits_k {{{\left\| {{{\bm{h}}_k}{{\bm{x}}^{\cal Q}_k}} \right\|}^2}} /\sigma _n^2$.

Subsequently, the transmission signals are recovered by the linear minimum mean-squared error (L-MMSE) detector with the estimated channel state information (CSI),
\begin{equation}\label{mimo-detection}
    {\bf{\widehat X}} = {{\bf{\widehat H}}^H}{\left( {{\bf{\widehat H}}{{{\bf{\widehat H}}}^H} + \sigma_n^2 {\bf{I}}} \right)^{ - 1}}{\bf{Y}},
\end{equation}
where ${\bf{\widehat X}}^T=\left [\hat { \bm  x}^{\cal Q}_{1}; \hat {\bm x}^{\cal Q}_{2};\cdots;\hat {\bm x}^{\cal Q}_{K} \right ]\in {\mathbb{C} }^{L_C\times K}$ is the recovered transmission signals, $\widehat {\bf H} = {\bf H} + {\Delta {\bf H}} $ is the estimated CSI, in which ${\Delta {\bf H}}$ is the estimation error with  ${\Delta {\bf H}} \in {\cal CN}(0, \sigma^2_e)$.

The semantic information from the $k$-th user, ${\hat {\bm z}^{\cal Q}_k} \in \mathbb{R}^{L_S \times 1}$, is recovered by the  {JSC} decoder as
\begin{equation}
    {\hat {\bm z}^{\cal Q}_k} = {C^{ - 1}}\left( {{\hat {\bm x}^{\cal Q}_k};{\bm{\gamma}}^{\cal Q}_k } \right),
\end{equation}
where ${C^{ - 1}}\left( {{\hat {\bm x}^{\cal Q}_k};{\bm{\gamma}}^{\cal Q}_k } \right)$\footnote{In order to reduce the number of representation symbols, we use $\cdot^{-1}$ here to represent the decoder.} \textcolor{black}{is JSC decoder for the $k$-th user with the modality ${\cal Q}$} and the learned parameters ${\bm{\gamma}}^{\cal Q}_k$.  {The JSC decoder aims to decompress the semantic information while mitigating the effects of channel distortion and inter-user interference.} According to the independence of transmission semantic information, we will have the single-modal semantic receiver and the multimodal semantic  receiver. 

\subsubsection{Single-Modal Semantic Receiver}
For single-modal semantic transmission, the semantic information from each user is exploited to perform different tasks independently. The recovered semantic information is  {employed} for the task of the $k$-th user by
\begin{equation}
   {\bm p}^{\cal Q}_k  = S^{-1} \left( {{\hat {\bm z}^{\cal Q}_k};{\bm{\varphi}}^{\cal Q}_k} \right) ,
\end{equation}
where ${\bm p}^{\cal Q}_k$ is the result of the task, i.e., the translated sentence for the machine learning task, and retrieval results for the image retrieval task.  $S^{-1}(;{\bm{\varphi}}^{\cal Q}_k)$ is the modality $\cal Q$ semantic decoder for the $k$-th user with learning parameters ${\bm{\varphi}}^{\cal Q}_k$.

\subsubsection{Multimodal Semantic Receiver}
With the multimodal semantic information, the final task is performed directly by merging the semantic information from different users. This is expressed by
\begin{equation}\label{eq-5}
    {\bm p} = S^{-1} \left( {\hat{\bm z}^{\cal Q}_1, \hat {\bm z}^{\cal Q}_2, \cdots, \hat {\bm z}^{\cal Q}_{K};{\bm{\varphi}}_{(1,2,\cdots,K)} } \right),
\end{equation}
where ${\bm p}$ is the results of the multimodal task and $S^{-1}\left( {;{\bm{\varphi}}_{(1,2,\cdots,K)} } \right)$ is the multimodal semantic decoder with learnable parameters ${\bm{\varphi}}_{(1,2,\cdots,K)}$.

\begin{figure*}
    \centering
    \includegraphics[width=180mm]{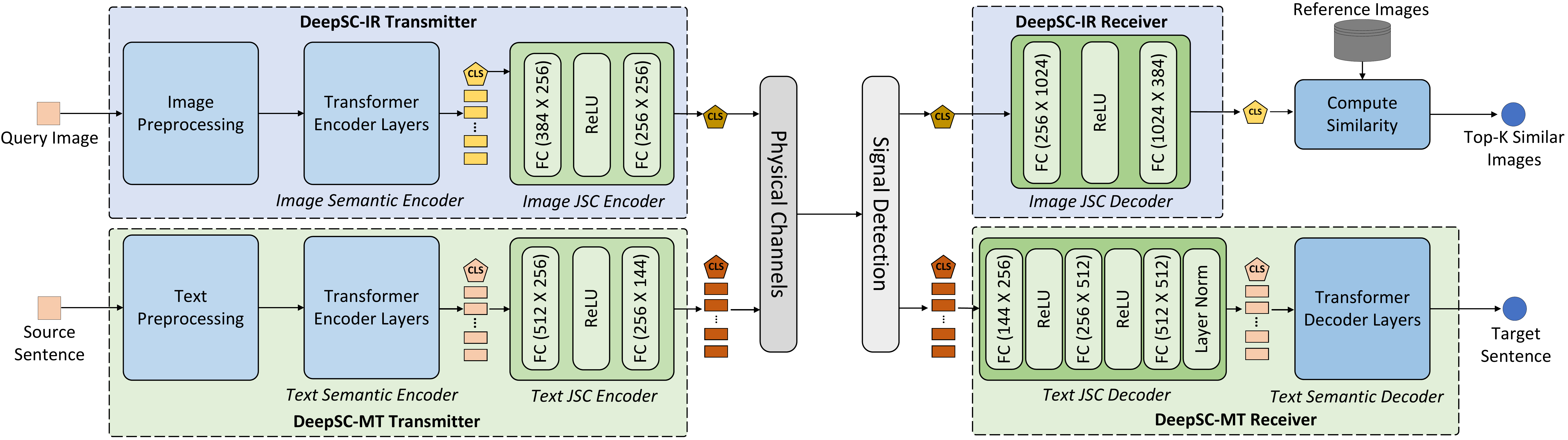}
    \caption{The network structure of single-modal multi-user semantic communications, which contains the DeepSC-IR transceiver and DeepSC-MT transceiver.}
    \label{fig:deep-independent}
\end{figure*}

\section{Single-modal Multi-user Semantic Communications}
In this section, we focus on the  multi-user semantic communication system to transmit single-modal data from multiple users.  We propose semantic communication systems for the image retrieval task (i.e., DeepSC-IR), and the machine translation task (i.e., DeepSC-MT). Particularly, we adopt the vision Transformer for image understanding and text Transformer for text understanding, in which the vision Transformer and text Transformer are assumed to have the same network structure. 

\subsection{Image Retrieval Task}
Assume that ${\cal D}^{\cal I}_k=\left \{ ({\bm s}^{\cal I}_{k,j}, {l}^{\cal I}_{k,j}) \right \}_{j=1}^{D}$ with size $D$ is the training image dataset for the $k$-th user, where ${\bm s}^{\cal I}_{k,j}$  and $l^{\cal I}_{k,j}$ are the $j$-th image and its corresponding label in ${\cal D}^{\cal I}_k$, respectively.  ${S_{\tt IR}\left( {;{{\bm{\alpha }}^{\cal I}_k}} \right)}$, $C_{\tt IR}\left( {;{{\bm{\beta }}^{\cal I}_k}} \right)$, and ${C_{\tt IR}^{ - 1}}\left( {{{\bf }};{\bm{\gamma}}^{\cal I}_k } \right)$ represent the semantic encoder,  {JSC} encoder, and  {JSC} decoder of the $i$-th user for the image retrieval task, respectively. 

\subsubsection{Model Description}

The proposed image retrieval network is shown in Fig.~\ref{fig:deep-independent}. Specifically, the DeepSC-IR transmitter consists of an image semantic encoder to extract image semantic information to be transmitted and a  {JSC} encoder to compress the semantic information, where the semantic encoder includes multiple vision Transformer layers and the  {JSC} encoder uses dense layers with different units. Especially, we choose only the $<$CLS$>$ vector-token to be transmitted as it represents the global image information.  After transmitting and performing signal detection, the DeepSC-IR receiver employs the  {JSC} decoder with different units to recover the transmitted image semantic information. 

The recovered semantic information after the  {JSC} decoder at the receiver can be used to match the other image semantic information in the database by computing the euclidean distance to find similar images as 
\begin{equation}
    d({\bm z}^{\cal I}_{k,j}, {\bm z}^{\cal I}_{k,i}) = \left \| {\bm z}^{\cal I}_{k,j} -  {\bm z}^{\cal I}_{k,i}\right \|_2.
\end{equation}
The euclidean distance becomes the cosine similarity when ${\bm z}^{\cal I}_{k,j}$ and ${\bm z}^{\cal I}_{k,i}$ are ${l}^2$ normalized.

\subsubsection{Training Algorithm}
As shown in Algorithm \ref{algorithm-1}, the training process of the DeepSC-IR consists of two phases due to different loss functions. The first phase is to train the semantic encoder, and the second phase is to train the  {JSC} codec. 

In the first phase, the semantic encoder will be trained by the function, \texttt{Train Semantic Encoder}. Different from other tasks, image retrieval is performed by computing the distance between images to return  similar images. Therefore, we choose metric learning, as one type of self-supervised learning, as the learning paradigm. Such paradigm aims at minimizing the distance between images belonging to the same category and maximizing the distance between images belonging to different categories. The loss function is expressed by 
\begin{equation}\label{eq14}
\begin{aligned}
    {\cal L}_{\tt IR}=&{\mathbb E}{\left [\sum\limits_{{l}^{\cal I}_{k,j}{=l}^{\cal I}_{k,i}}{\left (1-({\bm z}^{\cal I}_{k,j})^{\text T}{\bm z}^{\cal I}_{k,i} \right )} \right ]}\\
    &+ {\mathbb E}{\left [\sum\limits_{{l}^{\cal I}_{k,j}{\ne l}^{\cal I}_{k,i}}{\left (({\bm z}^{\cal I}_{k,j})^{\text T}{\bm z}^{\cal I}_{k,i}-\xi  \right )_{+}} \right ]},
\end{aligned}
\end{equation}
where the operator $(x)_+$ returns $\text{max}\left(x,0 \right)$, ${\bm z}_{i,j}$ is the image semantic information, $\xi$ is a constant margin to prevent the training signal from being overwhelmed by easy negatives. After training the semantic encoder with \eqref{eq14}, the semantic encoder becomes capable of extracting semantic image information, which returns a smaller euclidean distance if they are from images within the same category. 

In order to compress semantic redundancy while overcoming the distortion from the channels, the  {JSC} codec is trained in the second phase. The mean-squared error (MSE) is employed as the loss function to minimize the difference between the transmitted and recovered semantic image information, which is represented as
\begin{equation}\label{eq15}
    {\cal L}_{\tt MSE}= {\mathbb E}\left [ \left \|{ \hat {\bm  z}}^{\cal I}_{k,j} - {\bm z}^{\cal I}_{k,j} \right \|^2_2 \right ],
\end{equation}
where ${ \hat {\bm  z}}^{\cal I}_{k,j}$ is the semantic image information recovered at receiver and ${ {\bm  z}}^{\cal I}_{k,j}$ is the transmitted semantic image information. By minimizing the ${\cal L}_{\tt MSE}$,  {the JSC codec will learn to compress and decompress semantic image information for fewer transmitted symbols while keeping the semantic recovery accurately by dealing with the distortion and interference jointly from the channels and inter-users.}

\begin{algorithm}[!t]
\caption{DeepSC-IR Training Algorithm.}
\label{algorithm-1}
\SetKwInput{KwInput}{Input}                
\SetKwInput{KwInitia}{Initialization}                
\SetKwInput{KwOutput}{Output}              
\SetKwInput{KwRet}{Return}
\DontPrintSemicolon
  
  \KwInitia{The training dataset ${\cal D}^{\cal I}_k$ and the batch size $B$. } 

  \SetKwFunction{FMain}{Main}
  \SetKwFunction{FSE}{Train Semantic Encoder}
  \SetKwFunction{FCC}{Train  {JSC} Codec}
  \SetKwFunction{FWN}{Train Whole Network}
  \SetKwProg{Fn}{Function}{:}{}
  \Fn{\FSE{}}{
  \KwInput{Choose mini-batch data $\left \{ ({\bm s}^{\cal I}_{k, j}, l^{\cal I}_{k, j}) \right\}_{j = n}^{n + B} $ from ${\cal D}^{\cal I}_k$.}
   		$\left \{ {S_{\tt IR}\left( {{\bm s}^{\cal I}_{k,j};{{\bm{\alpha }}^{\cal I}_k}} \right)} \right\}_{j = n}^{n + B} \to \left \{ {\bm z}^{\cal I}_{k,j} \right\}_{j = n}^{n + B}$, \; 
   		Compute the ${\cal L}_{\tt IR}$ by \eqref{eq14} with $\left \{ {\bm z}^{\cal I}_{k,j} \right\}_{j = n}^{n + B}$,\;
        Train ${\bm{\alpha }}^{\cal I}_k$ $\to$ Gradient descent with ${\cal L}_{\tt IR}$,\;
        \KwRet{${S_{\tt IR}\left( {;{{\bm{\alpha }}^{\cal I}_k}}\right)}$.} 
  }

  
  \SetKwProg{Fn}{Function}{:}{}
  \Fn{\FCC{}}{
        \KwInput{The semantic image information $\left \{ {\bm z}^{\cal I}_{k,j}\right\}_{j = n}^{n + B}$.}
        \For{$j = n \to n + B$}{
        \textbf{Transmitter}:\;
   		\quad $ {C_{\tt IR}\left( {{\bm z}^{\cal I}_{k,j};{{\bm{\beta }}^{\cal I}_k}} \right)}  \to  {\bm x}^{\cal I}_{k,j} $, \; 
   		\quad Transmit ${\bm x}^{\cal I}_{k,j} $  over the channel,\;
   		\textbf{Receiver}:\;
   		\quad Receive ${\bf Y}$, \;
   		\quad MIMO detection by \eqref{mimo-detection} to get ${\hat {\bm x}}^{\cal I}_{k,j}$,\;
   		\quad $ {C^{-1}_{\tt IR}\left( {\hat {\bm x}^{\cal I}_{k,j};{{\bm{\gamma }}^{\cal I}_k}} \right)} \to  {\hat {\bm z}}^{\cal I}_{k,j} $, \;}
   		Compute the ${\cal L}_{\tt MSE}$ by \eqref{eq15} with ${{\bm z}}^{\cal I}_{k,j}$, ${\hat {\bm z}}^{\cal I}_{k,j}$,\;
        Train ${{\bm{\beta }}^{\cal I}_k}, {{\bm{\gamma }}^{\cal I}_k}$ $\to$ Gradient descent with $ {\cal L}_{\tt MSE}$,\;
        \KwRet{ ${C_{\tt IR}\left( {;{{\bm{\beta }}^{\cal I}_k}} \right)}, {C^{-1}_{\tt IR}\left( {;{{\bm{\gamma }}^{\cal I}_k}} \right)}$.}
  }
\end{algorithm}

\begin{algorithm}[!t]
\caption{DeepSC-MT Training Algorithm.}
\label{alg-2}
\SetKwInput{KwInput}{Input}                
\SetKwInput{KwInitia}{Initialization}
\SetKwInput{KwOutput}{Output}              
\SetKwInput{KwRet}{Return}
\DontPrintSemicolon
  
  \KwInitia{The training dataset ${\cal D}^{\cal T}_k$ and the batch size $B$.}

  \SetKwFunction{FMain}{Main}
  \SetKwFunction{FSE}{Train Semantic Codec}
  \SetKwFunction{FCC}{Train  {JSC} Codec}
  \SetKwFunction{FWN}{Train Whole Network}
  \SetKwProg{Fn}{Function}{:}{}
  \Fn{\FSE{}}{
        \KwInput{Choose mini-batch data $\left \{ ({\bm s}^{\cal T}_{k, j}, {\bm p}^{\cal T}_{k, j}) \right\}_{j = n}^{n + B} $ from ${\cal D}^{\cal T}_k$.}
   	    \For{$j = n \to n + B$}{
   		$ {S_{\tt MT}\left( {{\bm s}^{\cal T}_{k,j};{{\bm{\alpha }}^{\cal T}_k}} \right)}  \to  {\bm z}^{\cal T}_{k, j} $, \; 
   		$ {S^{-1}_{\tt MT}\left( {{\bm z}^{\cal T}_{k,j};{\bm{\varphi}}^{\cal T}_k} \right)}  \to  \hat {\bm p}^{\cal T}_{k, j} $, \; }
   		Compute  ${\cal L}_{\tt MT}$ by \eqref{loss-ce} with  ${\bm p}^{\cal T}_{k, j}$ and $\hat {\bm p}^{\cal T}_{k, j}$.\;
        Train ${\bm{\alpha }}^{\cal T}_k, {\bm{\varphi}}^{\cal T}_k$ $\to$ Gradient descent with ${\cal L}_{\tt MT}$.\;
        \KwRet{${S_{\tt MT}\left( {;{{\bm{\alpha }}^{\cal T}_k}}\right)}$ and ${S^{-1}_{\tt MT}\left( {;{{\bm{\varphi }}^{\cal T}_k}}\right)}$.} 
  }

  
  \SetKwProg{Fn}{Function}{:}{}
  \Fn{\FCC{}}{
       \KwInput{The semantic text features  $\left \{ {\bm z}^{\cal T}_{k,j} \right\}_{j = n}^{n + B}$.}
        \For{$j = n \to n + B$}{
        \textbf{Transmitter}:\;
   		\quad $ {C_{\tt MT}\left( {{\bm z}^{\cal T}_{k,j};{{\bm{\beta }}^{\cal T}_k}} \right)}  \to  {\bm x}^{\cal T}_{k,j} $, \; 
   		\quad Transmit ${\bm x}^{\cal T}_{k,j} $  over the channel.\;
   		\textbf{Receiver}:\;
   		\quad Receive ${\bf Y}$, \;
   		\quad MIMO detection by \eqref{mimo-detection} to get ${\hat {\bm x}}^{\cal T}_{k,j}$,\;
   		\quad $ {C^{-1}_{\tt MT}\left( {\hat {\bm x}^{\cal T}_{k,j};{{\bm{\gamma }}^{\cal T}_k}} \right)} \to  {\hat {\bm z}}^{\cal T}_{k,j} $, \; }
   		Compute  ${\cal L}_{\tt MSE}$ with \eqref{mt-mse}.\;
        Train ${{\bm{\beta }}^{\cal T}_k}, {{\bm{\gamma }}^{\cal T}_k}$ $\to$ Gradient descent with ${\cal L}_{\tt MSE}$.\;
        \KwRet{${C_{\tt MT}\left( {;{{\bm{\beta }}^{\cal T}_k}} \right)}$ and ${C^{-1}_{\tt MT}\left( {;{{\bm{\gamma }}^{\cal T}_k}} \right)}$.}
  }

  \SetKwProg{Fn}{Function}{:}{}
  \Fn{\FWN{}}{
        
         \KwInput{Choose mini-batch data $\left \{ ({\bm s}^{\cal T}_{k, j}, {\bm p}^{\cal T}_{k, j}) \right\}_{j = n}^{n + B} $ from ${\cal D}^{\cal T}_k$.}
        \For{$j = n \to n + B$}{
        Repeat line 3-4, 11-16, and 4 to get ${\hat {\bm p}}^{\cal T}_{k, j}$,\;}
  		Compute ${\cal L}_{\tt MT}$ by \eqref{loss-ce} with ${\bm p}^{\cal T}_{k,j}$ and ${\hat {\bm p}}^{\cal T}_{k, j}$. \;
        Train ${{\bm{\alpha }}^{\cal T}_k}, {{\bm{\beta }}^{\cal T}_k}, {{\bm{\gamma }}^{\cal T}_k}, {{\bm{\varphi }}^{\cal T}_k}$ $\to$ Gradient descent with ${\cal L}_{\tt MT}$.\;
        \KwRet{${S_{\tt MT}\left( {;{{\bm{\alpha }}^{\cal T}_k}}\right)}$, ${S^{-1}_{\tt MT}\left( {;{{\bm{\varphi }}^{\cal T}_k}}\right)}$, ${C_{\tt MT}\left( {;{{\bm{\beta }}^{\cal T}_k}} \right)}$, and ${C^{-1}_{\tt MT}\left( {;{{\bm{\gamma }}^{\cal T}_k}} \right)}$. }
  }
\end{algorithm}

\subsection{Machine Translation Task}
Assume ${\cal D}^{\cal T}_k=\left \{ ({\bm s}^{\cal T}_{k,j}, {\bm p}^{\cal T}_{k,j}) \right \}_{j=1}^{D}$ with size $D$ as the training text dataset for the $k$-th user, where ${\bm s}^{\cal T}_{k,j}$  and ${\bm p}^{\cal T}_{k,j}$ are the $j$-th sentence in the source language and the translated sentence in the target language, respectively. ${\bm s}^{\cal T}_{k,j}[n]$ and ${\bm p}^{\cal T}_{k,j}[n]$ represent the $n$-th word in sentence ${\bm s}^{\cal T}_{k,j}$  and ${\bm p}^{\cal T}_{k,j}$, respectively. ${S_{\tt MT}\left( {;{{\bm{\alpha }}^{\cal T}_k}} \right)}$, $C_{\tt MT}\left( {;{{\bm{\beta }}^{\cal T}_k}} \right)$, ${C_{\tt MT}^{ - 1}}\left( {{{\bf }};{\bm{\gamma}}^{\cal T}_k} \right)$, and ${S^{-1}_{\tt MT}\left( {;{{\bm{\varphi}}^{\cal T}_k}} \right)}$ represent the semantic encoder,  {JSC} encoder,  {JSC} decoder, and semantic decoder of the $k$-th user for the machine translation task, respectively.

\subsubsection{Model Description}
The proposed machine translation network is shown in Fig. \ref{fig:deep-independent}. The transmitter includes a text semantic encoder and a text  {JSC} encoder to extract and compress the semantic text information, respectively, where the text semantic encoder adopts multiple Transformer encoder layers  and the designed text  {JSC} encoder  in Fig.~\ref{fig:deep-independent} is with multiple dense layers. At the receiver, the designed text  {JSC} decoder recovers the semantic text information from distorted signals. Subsequently, the semantic decoder consists of multiple Transformer decoder layers to derive the translated sentence based on the recovered semantic text information.

\subsubsection{Training Algorithm}
As shown in Algorithm \ref{alg-2}, the training process of DeepSC-MT consists of three phases: \texttt{Train Semantic Codec}, \texttt{Train  {JSC} Codec}, and \texttt{Train Whole Network}. 

The first phase is \texttt{Train Semantic Codec}. The semantic codec, ${S_{\tt MT}\left( {;{{\bm{\alpha }}^{\cal T}_k}}\right)}$ and ${S^{-1}_{\tt MT}\left( {;{{\bm{\varphi }}^{\cal T}_k}}\right)}$, will be trained firstly by the cross-entropy (CE) loss function, which enables the model to convert the meaning to the target sentence by learning the target language word distribution. The CE loss function is represented by
\begin{equation}\label{loss-ce}
     {\mathcal L}_{\tt MT}= {\mathbb E}\left [ -\sum_{n}{P({\bm p}^{\cal T}_{k, j}[n])}\text{log} \left (P(\hat {\bm p}^{\cal T}_{k, j}[n])\right) \right ],
\end{equation}
where $P(\hat {\bm p}^{\cal T}_{k, j}[n])$ is the predicted probability
that the $n$-th word appears in sentence $\hat {\bm p}^{\cal T}_{k, j}$, and $P({\bm p}^{\cal T}_{k, j}[n])$ is the real probability that the $n$-th word appears in the sentence ${\bm p}^{\cal T}_{k, j}$. After convergence, the model learns the syntax, phrase, the meaning of words in the target language. 

In the second training phase that is listed as \texttt{Train  {JSC} Codec} of Algorithm 2,  {the JSC codec, $C_{\tt MT}(; \bm{\beta}^{\cal T}_k)$ and $C^{-1}_{\tt MT}(; \bm{\gamma}^{\cal T}_k)$, are also trained to learn the compress and decompress semantic text information, as well as deal with the channel distortion and multi-user interference} with the MSE loss function given by
\begin{equation}\label{mt-mse}
    {\cal L}_{\tt MSE}= {\mathbb E}\left [ \left \|{ \hat {\bm  z}}^{\cal T}_{k,j} - {\bm z}^{\cal T}_{k,j} \right \|^2_2 \right ],
\end{equation}
where ${ \hat {\bm  z}}^{\cal T}_{k,j}$ is the recovered semantic text information at the receiver and ${ {\bm  z}}^{\cal T}_{k,j}$ is the transmitted semantic text information.

Different from the DeepSC-IR training algorithm, there exists a semantic decoder at the DeepSC-MT receiver. This means that semantic errors between ${ \hat {\bm  z}}^{\cal T}_{k,j}$ and ${ {\bm  z}}^{\cal T}_{k,j}$ can be mitigated by jointly training the whole system shown as \texttt{Train Whole Network} in Algorithm 2 with the loss function \eqref{loss-ce}.

\begin{figure*}
    \centering
    \includegraphics[width=180mm]{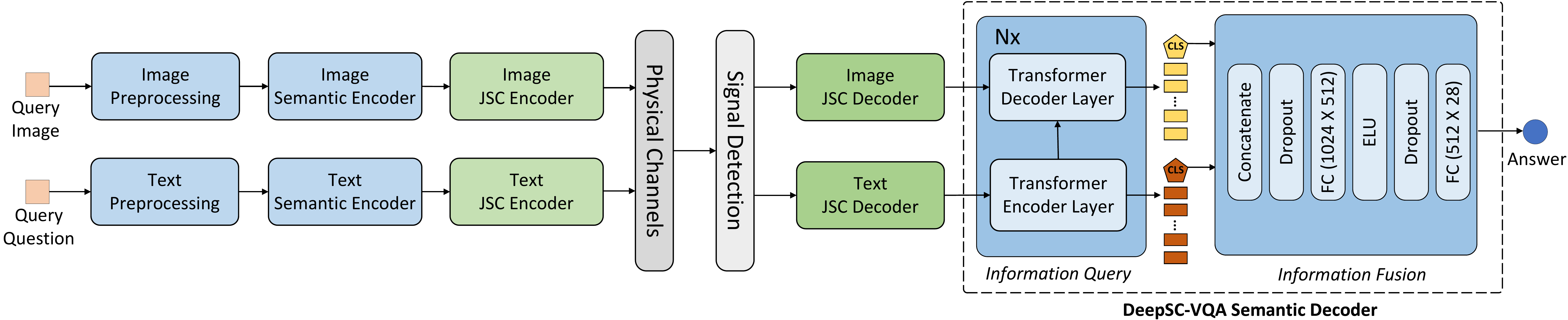}
    \caption{The proposed network structure of multimodal multi-user semantic communication system with DeepSC-VQA transceiver.}
    \label{fig:deep-dependent}
\end{figure*}

\section{Multimodal Multi-user Semantic Communications}
In this section, the multimodal multi-user semantic communications are investigated for serving the VQA task, namely DeepSC-VQA, in which the transmitters adopt the same structures as that of DeepSC-IR for images and DeepSC-MT for texts. They also share the same  {JSC} decoder design. Particularly, a novel semantic decoder is proposed to merge the image-text semantic information.

\subsection{Model Description}
Assume that the $k$-th user for image transmission and the $i$-th user for text transmission, ${\cal D}^{\cal I, T}_{k,i} = \left \{ ({\bm s}^{\cal I}_{k,j}, {\bm s}^{\cal T}_{i,j}, {l}_{(k,i),j}) \right \}_{j=1}^{D}$ with size $D$ is the training dataset, where ${\bm s}^{\cal I}_{k,j}$ is the $j$-th image from the $k$-th user, ${\bm s}^{\cal T}_{i,j}$ is the $j$-th text from the $i$-th user, and $l_{(k,i),j}$ is the answer label for ${\bm s}^{\cal I}_{k,j}$ and ${\bm s}^{\cal T}_{i,j}$. ${S_{\tt VQA}\left( {;{{\bm{\alpha }}^{\cal I}_k}} \right)}$, $C_{\tt VQA}\left( {;{{\bm{\beta }}^{\cal I}_k}} \right)$, ${C_{\tt VQA}^{ - 1}}\left( {{{\bf }};{\bm{\gamma}}^{\cal I}_k } \right)$ are the image semantic  encoder, image  {JSC} encoder, and image  {JSC} decoder of the $i$-th user, respectively. ${S_{\tt VQA}\left( {;{{\bm{\alpha }}^{\cal T}_i}} \right)}$, $C_{\tt VQA}\left( {;{{\bm{\beta }}^{\cal T}_i}} \right)$, ${C_{\tt VQA}^{ - 1}}\left( {{{\bf }};{\bm{\gamma}}^{\cal T}_i } \right)$ are the text semantic encoder, text  {JSC} encoder, and text  {JSC} decoder of the $k$-th user, respectively. ${S^{-1}_{\tt VQA}\left( {;{{\bm{\varphi}}_{(k,i)}}} \right)}$ represents joint semantic decoder of the $i$-th and the $t$-th user for the VQA task.

As shown in Fig. \ref{fig:deep-dependent}, the proposed DeepSC-VQA network consists of one image transmitter, one text transmitter, and one receiver for simplicity. For the DeepSC-VQA transmitters and receivers, we adopt the same structures as the image transmitter of DeepSC-IR and text transmitter of DeepSC-MT to unify the transmitter paradigm. At the receiver, the structures of the image  {JSC} decoder and text  {JSC} decoder are also the same as that of the image  {JSC} decoder in DeepSC-IR and that of the text  {JSC} decoder in DeepSC-MT. Besides, we develop a new semantic decoder network for image-text information fusion,  which includes two modules: information query module and information fusion module. 

\subsubsection{Information Query}
The layer-wise Transformer is adopted. Different from the classic Transformer, where the decoder layers exploit the output tokens of the last layer of encoder as the input, the layer-wise Transformer employs the output tokens of each encoder layer as the input of each decoder layer. Such a design can leak more text information than classic Transformer and guide the image information query in the decoder more efficiently, which does not introduce any costs.

\subsubsection{Information Fusion}
After the information query, the layer-wise Transformer has already captured keywords in the text information and the corresponding regions in image information, which has reflected in the output tokens. We will then need to fuse keywords and the corresponding image regions to get the answer. As mentioned in Section II, the $<$CLS$>$ token represents the global descriptor. Therefore, the $<$CLS$>$ tokens in the output tokens of the Transformer encoder and Transformer decoder represent the global text information and global image information, respectively. Using the text $<$CLS$>$ and image $<$CLS$>$, we design the information fusion module as shown in Fig. \ref{fig:deep-dependent}, where dropout layers are used here to avoid over-fitting.

\subsection{Training Algorithm}
Similar to the DeepSC-MT training algorithm, the DeepSC-VQA is trained jointly by three phases but with different loss functions. 

The first phase is \texttt{Train Semantic Codec},  {the semantic codec of DeepSC-VQA, ${S_{\tt VQA}\left( {;{{\bm{\alpha }}^{\cal I}_k}} \right)}$, ${S_{\tt VQA}\left( {;{{\bm{\alpha }}^{\cal T}_i}} \right)}$, ${S^{-1}_{\tt VQA}\left( {;{{\bm{\varphi }}_{(k,i)}}} \right)}$,} is trained jointly by the CE loss function, 
\begin{equation}\label{vqa-1}
    {\cal L}_{\tt VQA}= {\mathbb E}\left [ -{P\left(l_{(k,i), j} \right)}\text{log} \left (P\left(\hat l_{(k,i), j}\right)\right) \right ],
\end{equation}
where ${P(l_{(k,i), j}})$ and  ${P(\hat l_{(k,i), j}})$ are the real and predicted probability of answer, respectively. By reducing the loss value of CE, the network learns to predict the answer with the highest probability of accuracy.

After training the semantic codec,  {JSC} codecs are trained to compress by  {JSC encoder to reduce the number of transmitted symbols, and then decompress by the  {JSC} decoder to recover semantic information accurately over multiple user physical channels.} The image and text  {JSC} codec are trained jointly by the function \texttt{Train  {JSC} Codec}, in which the loss function is designed as
\begin{equation}\label{vqa-2}
    {\cal L}^{\tt{(VQA)}}_{\tt MSE}= {\mathbb E}\left [\left \|{ \hat {\bm  z}}^{\cal I}_{k,j} - {\bm z}^{\cal I}_{k,j} \right \|^2_2 + \left \|{ \hat {\bm  z}}^{\cal T}_{i,j} - {\bm z}^{\cal T}_{i,j} \right \|^2_2 \right ],
\end{equation}
where ${{\bm  z}}^{\cal I}_{k,j}$ and ${{\bm  z}}^{\cal T}_{i,j}$ are the transmitted semantic image and text information, respectively. ${\hat {\bm  z}}^{\cal I}_{k,j}$ and ${\hat {\bm  z}}^{\cal T}_{i,j}$ are the recovered semantic image and text information at the receiver, respectively. 

There exists error propagation from the  {JSC} decoders to the semantic receiver because of the imperfect semantic information recovery in the low SNR regimes. Therefore, the whole DeepSC-VQA network is trained jointly with loss function \eqref{vqa-1} to reduce the error propagation, which is the function \texttt{Train Whole Network}.

\section{Simulation Results}
In this section, we compare the proposed multi-user semantic communication systems with traditional source coding and channel coding methods over various channels, in which both the perfect and imperfect CSI are considered. 

\subsection{Implementation Details}
\subsubsection{The Datasets}
We choose four popular datasets commonly used for the image retrieval task. \textit{Stanford Online Products}~\cite{SongXJS16} consists of 120,053 online products images representing 22,634 categories, in which 11,318 categories are used for training and the remaining 11,316 categories are used for testing.  \textit{CUB-200-2011}~\cite{wah2011caltech} has 200 bird categories with 11,789 images. We split the first 100 classes for training and the rest of 100 classes for testing. \textit{Cars196}~\cite{Krause0DF13} contains 16,185 images corresponding to 196 car categories with the first 98 categories to be used for training. The remaining 98 categories are used for testing. \textit{In-Shop Clothes}~\cite{LiuLQWT16} contains 72,172 cloth images belonging to 7,986 categories, in which 3997 categories are used for training and the other 3985 categories will be used for testing.

For the machine translation task, we adopt the \textit{WMT 2018 Chinese-English news track}, which contains 202,221  pairs for training and 50,556 pairs for testing. The dataset is filtered into the length of English sentences with 5 to 75 words.

For the VQA task, we adopt the popular dataset: \textit{CLEVR}~\cite{JohnsonHMFZG17}, which consists of a training set of 70,000 images and 699,989 questions and a test set of 15,000 images and 149,991 questions.

\subsubsection{Training Settings}

The image semantic encoder of DeepSC-IR is based on the public implementation of DeiT-small model\footnote{https://github.com/facebookresearch/deit.} with 6 Transformer encoder layers. The setting of the \texttt{Train Semantic Encoder} of DeepSC-IR is the Adam optimizer with learning rate $3\times 10^{-5}$, weight decay $5\times 10^{-4}$, batch size of 64, and epoch of 40. The setting of the \texttt{Train  {JSC} Encoder} of DeepSC-IR is the Adam optimizer with learning rate $1\times 10^{-3}$, batch size of 64, and epoch of 100. During the training phase, the data augmentation is used to resize the image to 256 $\times$ 256 and then take a random crop of size 224 $\times$ 224 combined with random horizontal flipping. In the test phase, the images are resized to 256 $\times$ 256 first and centrally cropped to 224 $\times$ 224. 

The text semantic codec of DeepSC-MT is based on the public implementation of the Transformer model\footnote{https://huggingface.co/Helsinki-NLP.} with 6 Transformer encoder layers and decoder layers. The setting of the \texttt{Train Semantic Codec} of DeepSC-MT is the Adam optimizer with learning rate $1\times 10^{-5}$, betas of 0.9 and 0.98, batch size of 64, and epoch of 10. The setting of the \texttt{Train  {JSC} Codec} of DeepSC-MT is the Adam optimizer with learning rate $1\times 10^{-3}$,  batch size of 64, and epoch of 20. The setting of the \texttt{Train Whole Network} of DeepSC-MT is the same as that of \texttt{Train Semantic Codec} but with epoch of 20.

The image semantic encoder of DeepSC-VQA is also based on the pre-trained DeiT-small model but the other parts are trained from scratch, where the text semantic encoder is with 6 Transformer encoder layers and the semantic decoder is with 4 Transformer encoder layers and decoder layers. \textcolor{black}{We freeze the image semantic encoder to speed up training. The output dimension for the vision Transformer and text Transformer are set differently, which requires the dimension increasing operations after the image JSC decoder. The dimension-increasing operations successively include the dropout layer, dense layer from 384 to 512, ELU activation layer, dropout layer, and dense layer from 512 to 512, and  ELU activation layer.} The setting of the \texttt{Train Semantic Codec} of DeepSC-VQA is the Adam optimizer with learning rate $1\times 10^{-4}$, betas of 0.9 and 0.98, batch size of 64, and epoch of 80. The setting of the \texttt{Train JSC Codec} of DeepSC-VQA is the Adam optimizer with learning rate $1\times 10^{-3}$,  batch size of 128, and epoch of 30. The setting of \texttt{Train Whole Network} of DeepSC-MT is the same as that of the \texttt{Train Semantic Codec} but with epoch of 10. The data augmentation is used to resize images to 224 $\times$ 224 with BICUBIC interpolation for both training and testing.

\subsubsection{Benchmarks and Performance Metrics}
Our benchmark will adopt several typical source and channel coding methods. 
\begin{itemize}
    \item Error-free Transmission: The full, noiseless images and texts are delivered to the receiver, which will serve as the upper bound.
    \item Traditional Methods: To perform the source and channel coding separately, we use the following technologies, respectively:
    \begin{itemize}
        \item 8-bit unicode transformation format (UTF-8) encoding for text source coding, a commonly used method in text compression;
        \item Joint photographic experts group (JEPG) for image source coding, a widely used method in image compression;
        \item Turbo coding for text channel coding, popular channel coding for a small size file;
        \item Low-density parity-check code (LDPC) for image channel coding, and classic channel coding for big size files.
    \end{itemize}
\end{itemize} 
In the simulation, all coding rates of channel codings are 1/3. Perfect and imperfect CSI are set with $\sigma^2_e=0$  and $\sigma^2_e=0.025$, respectively.  {We set $r=2$ for Rician channels and ${\bf H} = \bf I$ for AWGN channels.}

The Recall@1 evaluation metric is adopted as performance metric for the image retrieval task. Bi-lingual evaluation understudy (BLEU) score is adopted for the machine translation task. Answer accuracy is used for VQA task.

\begin{figure*}[!t]
	\centering
	\hspace{-9mm}
	\subfigure[AWGN Channels.]{
			\includegraphics[width=60mm]{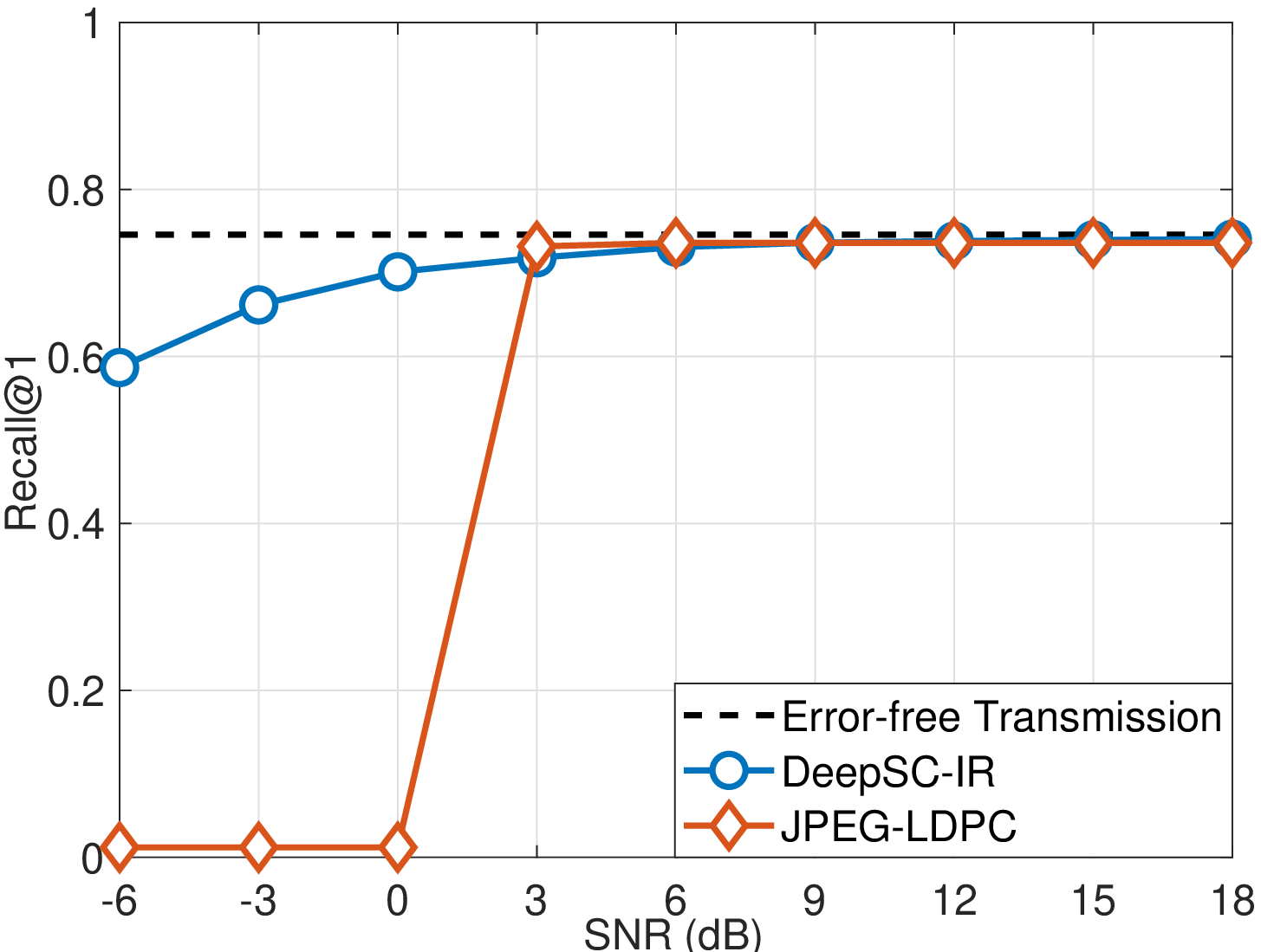} 
		\label{fig:cub200-1}
	}\hspace{-3mm}
    	\subfigure[Rayleigh Channels.]{
   		 	\includegraphics[width=60mm]{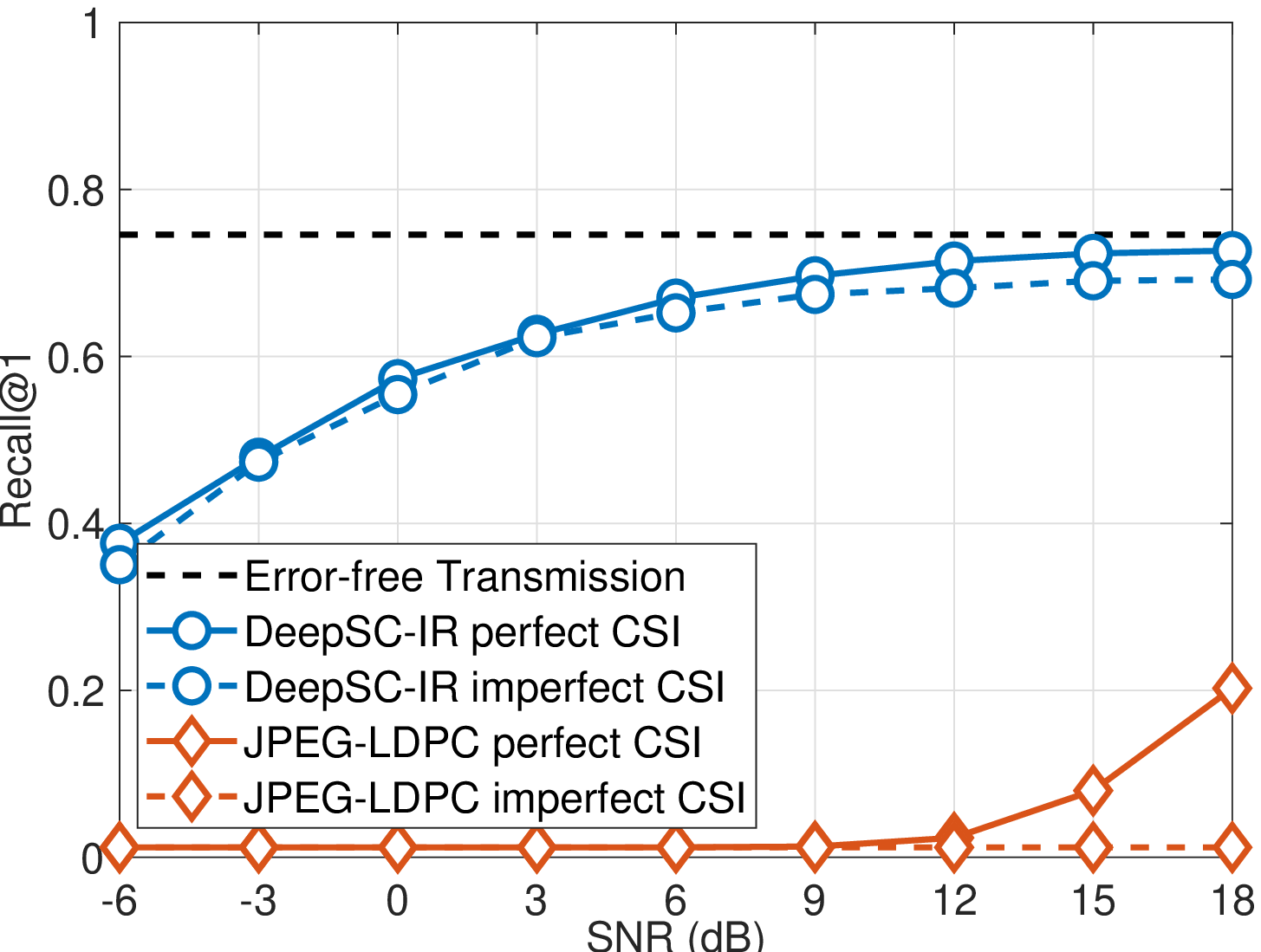}
		\label{fig:cub200-2}
    }\hspace{-3mm}
    	\subfigure[Rician Channels.]{
   		 	\includegraphics[width=60mm]{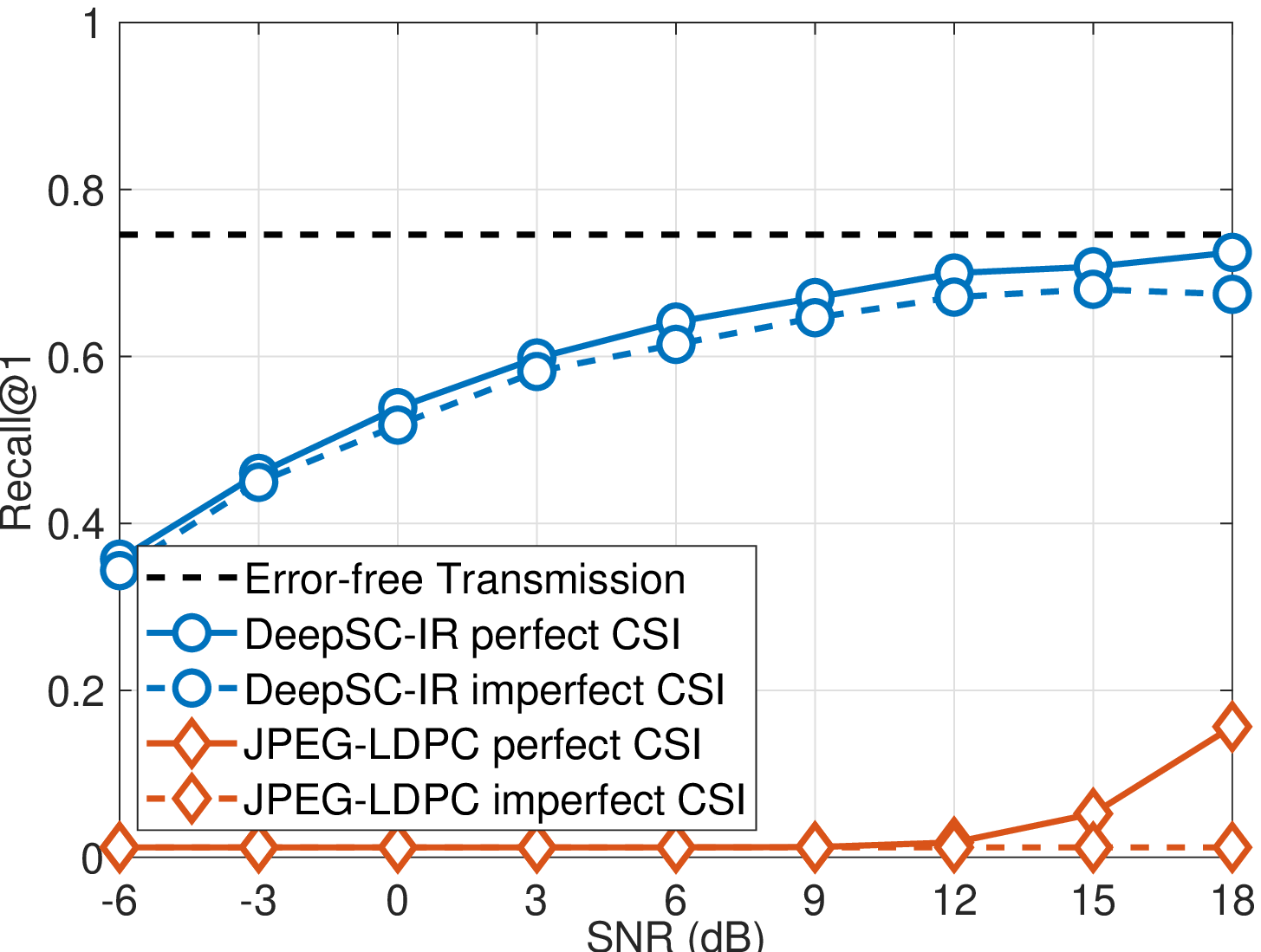}
		\label{fig:cub200-3}
    }\hspace{-9mm}
    \caption{Recall@1 comparison between DeepSC-IR and JPEG-LDPC with 8-QAM  over different channels, in which the dataset is CUB-200-2011.}
	\label{fig:cub200}
\end{figure*}
\begin{figure*}[!t]
	\centering
	\hspace{-9mm}
	\subfigure[Stanford Online Products.]{
			\includegraphics[width=60mm]{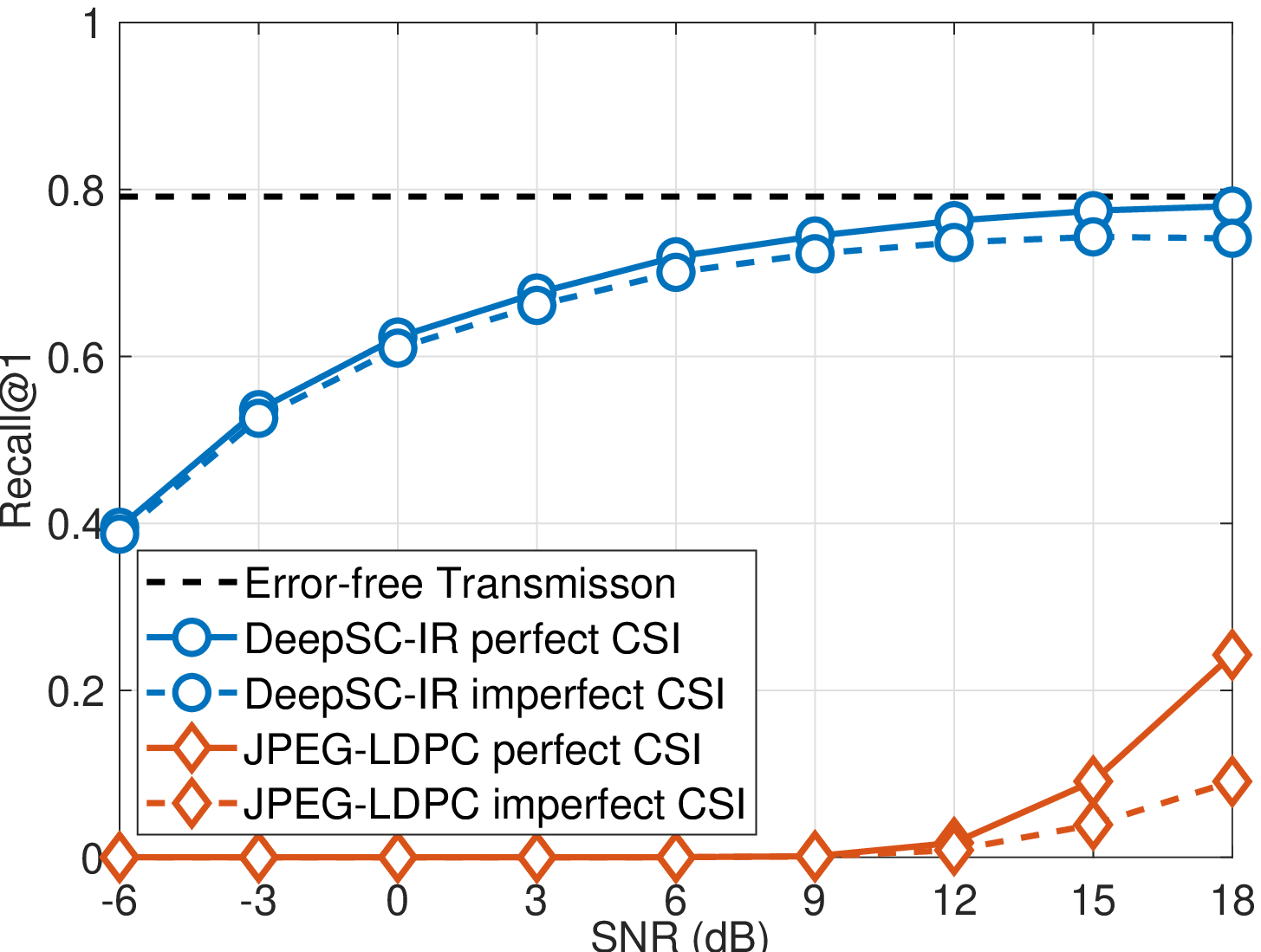} 
		\label{fig:diff-data-1}
	}\hspace{-3mm}
    	\subfigure[Cars196.]{
		 	\includegraphics[width=60mm]{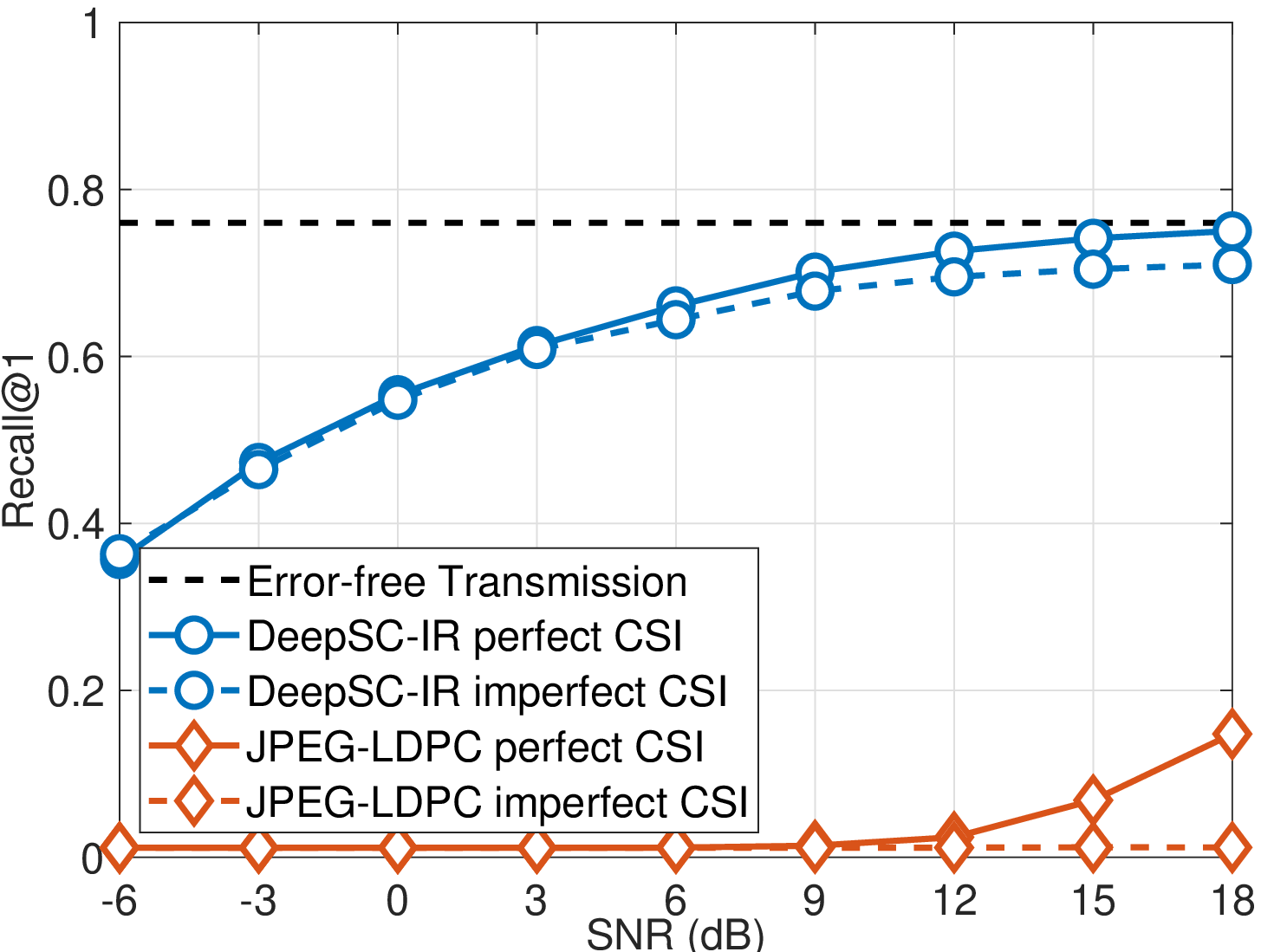}
		\label{fig:diff-data-2}
    }\hspace{-3mm}
    \subfigure[In-shop Clothes.]{
   		 	\includegraphics[width=60mm]{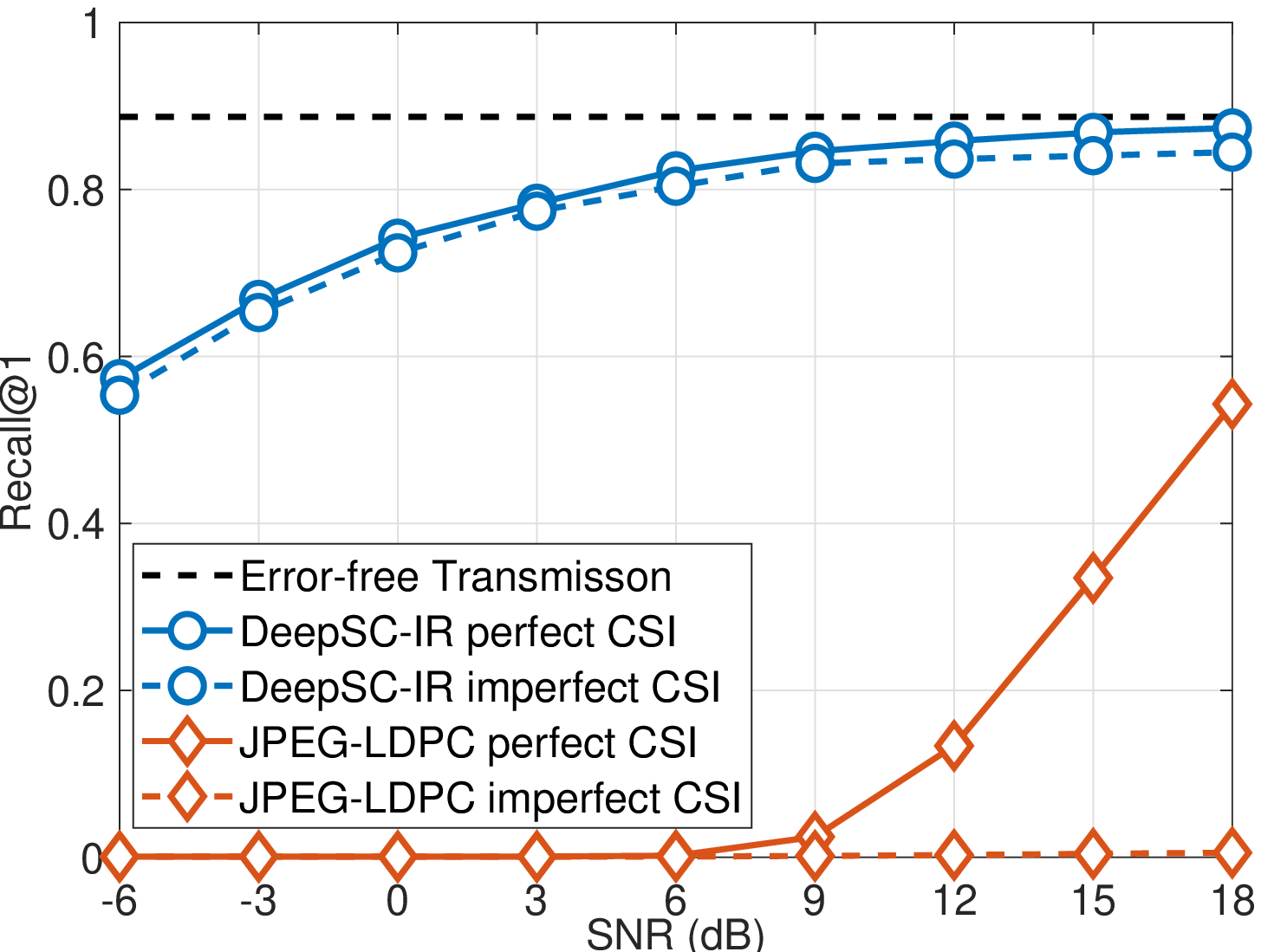}
		\label{fig:diff-data-3}
    }\hspace{-9mm}
	\caption{Recall@1 comparison between DeepSC-IR and JPEG-LDPC with 8-QAM for different datasets under Rician channels.}
	\label{fig:diff-data}
\end{figure*}

\begin{figure*}[!t]
	\centering
	\hspace{-9mm}
	\subfigure[AWGN Channels.]{
			\includegraphics[width=60mm]{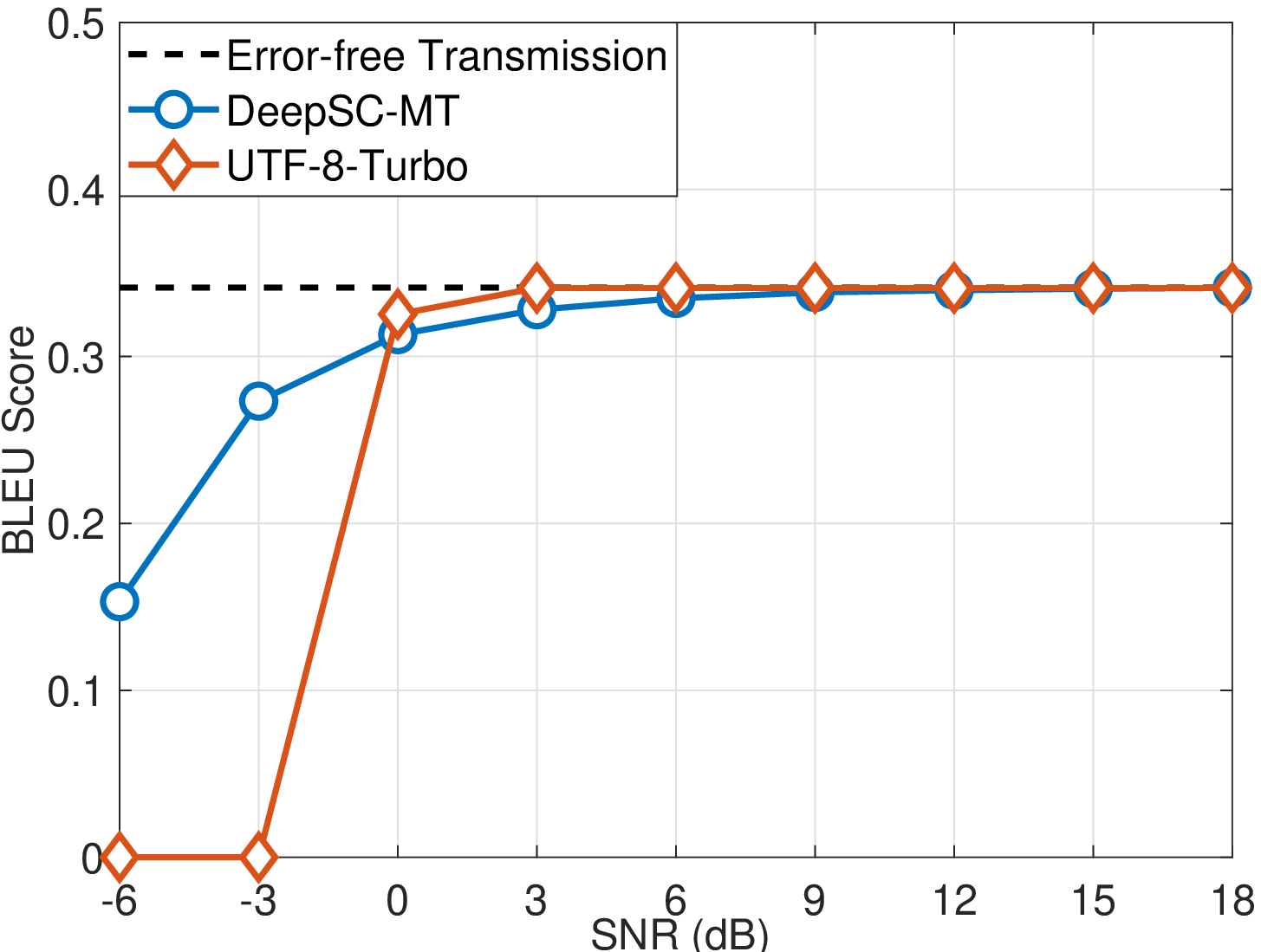}
		\label{fig:en-zh-1}
	}\hspace{-3mm}
    	\subfigure[Rayleigh Channels.]{
		 	\includegraphics[width=60mm]{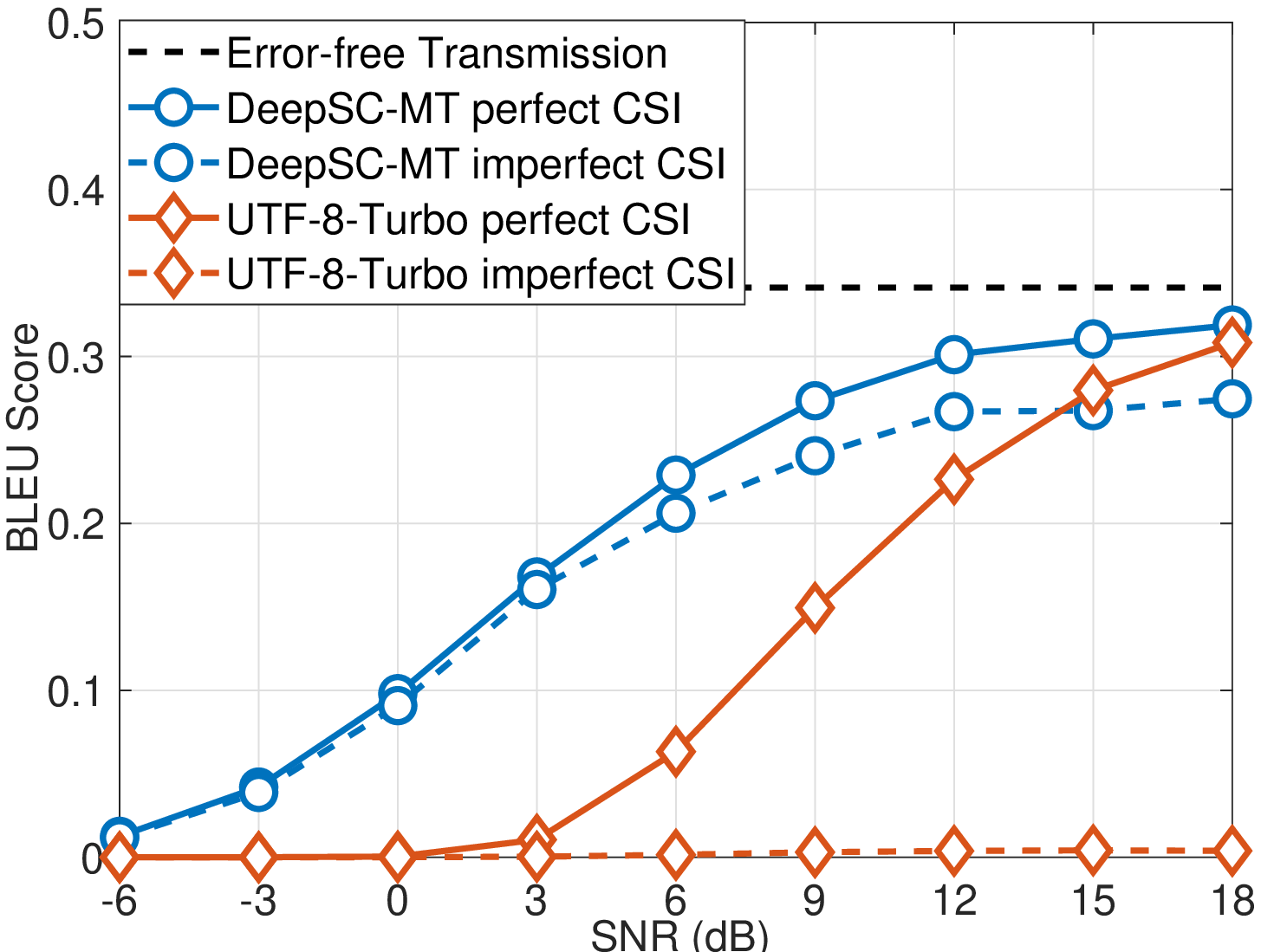}
		\label{fig:en-zh-2}
    }\hspace{-3mm}
    \subfigure[Rician Channels.]{
   		 	\includegraphics[width=60mm]{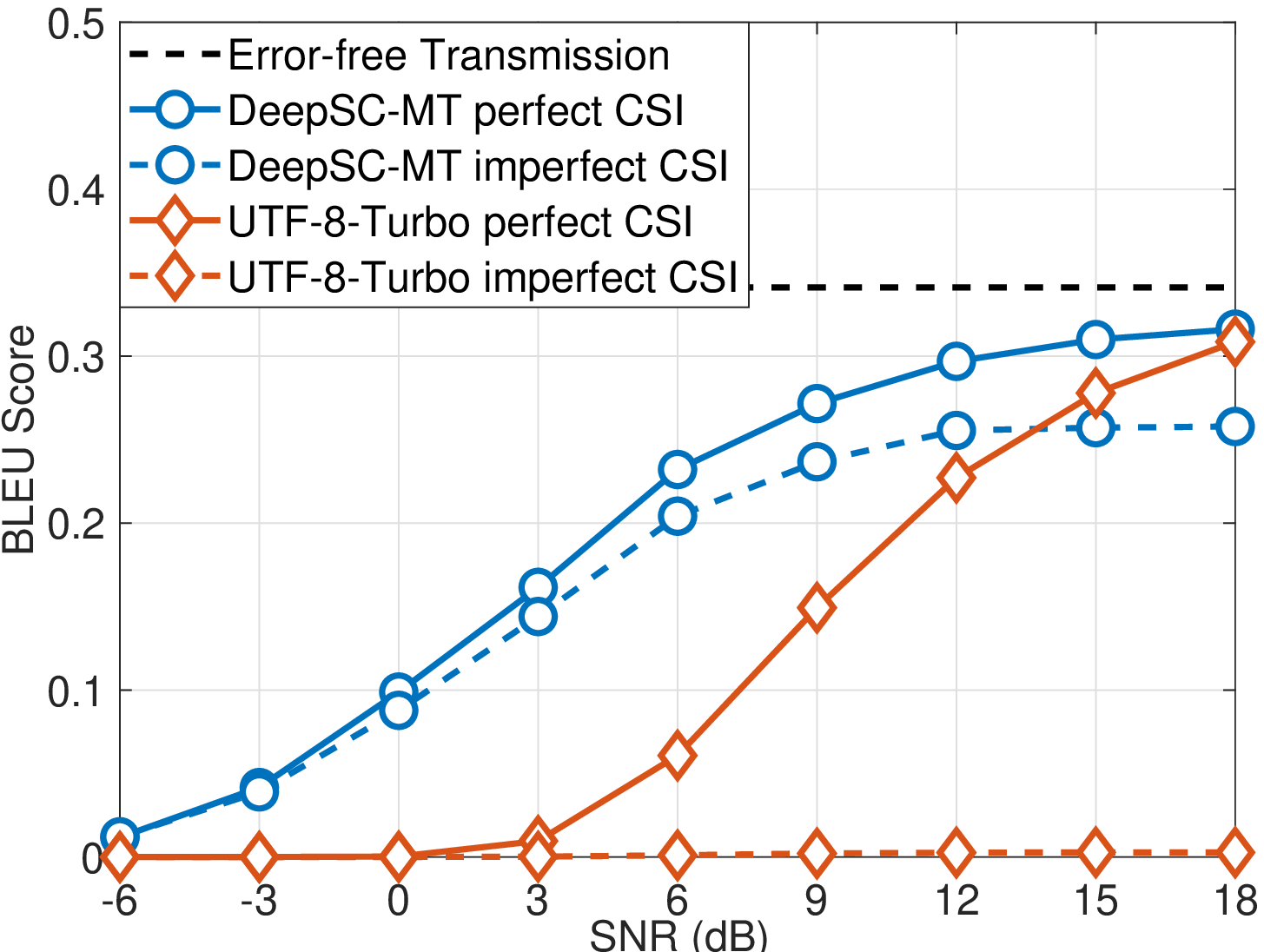}
		\label{fig:en-zh-3}
    }\hspace{-9mm}
	\caption{BLEU score comparison between DeepSC-MT and UTF-8-Turbo with QPSK for English-to-Chinese under  {AWGN channels, Rayleigh channels, and Rician channels.}}
	\label{fig:en-zh}
\end{figure*}

\subsection{Single-Modal Multi-User Semantic Communication}
The Recall@1 performance comparison for different channels on CUB-200-2011 and for different datasets over Rician channels are shown in Fig. \ref{fig:cub200} and Fig. \ref{fig:diff-data}, respectively. From Fig.~\ref{fig:cub200}, for different channels on CUB-200-2011, the proposed DeepSC-IR provides a significant gain at the low SNR regimes and approaches to the upper bound at the high SNR regimes among the reported methods, outperforming the JPEG-LDPC with 8-QAM by a margin of more than 24dB gain for 0.4 Recall@1 over fading channels. Even when using imperfect CSI, the DeepSC-IR still outperforms the benchmarks with slight performance degradation at Recall@1. From Fig.~\ref{fig:diff-data}, for different datasets over Rician channels, the DeepSC-IR also outperforms the JPEG-LDPC with 8-QAM in the three popular datasets at Recall@1 with more than 24 dB gain, respectively. Besides, exploiting imperfect CSI considerably decreases the performance at Recall@1 for the traditional method, especially in In-Shop Clothes but is only with a slightly performance degradation for DeepSC-IR.

\begin{figure*}[!t]
	\centering
	\hspace{-9mm}
	\subfigure[AWGN Channels.]{
			\includegraphics[width=60mm]{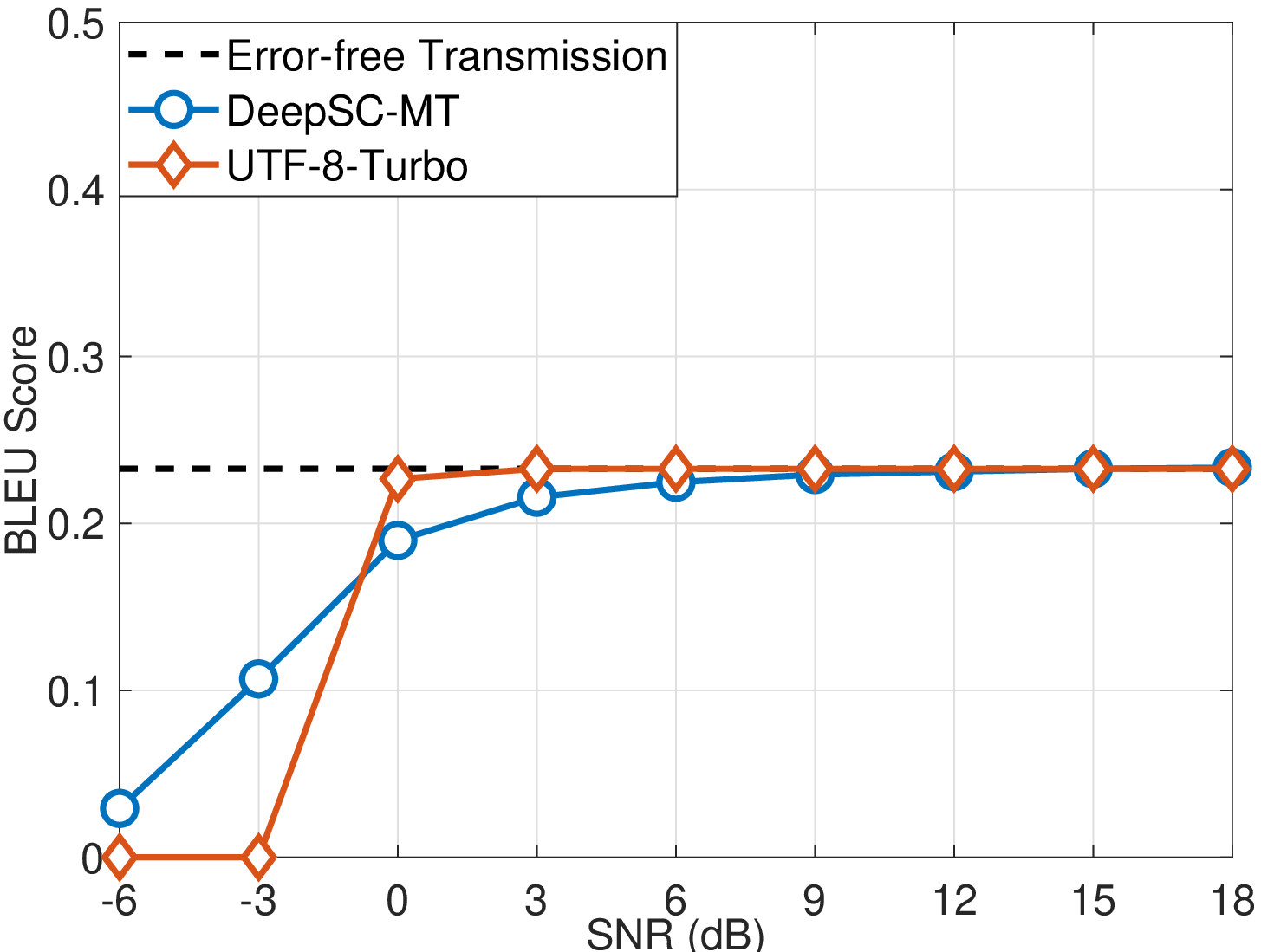}
		\label{fig:zh-en-1}
	}\hspace{-3mm}
    	\subfigure[Rayleigh Channels.]{
		 	\includegraphics[width=60mm]{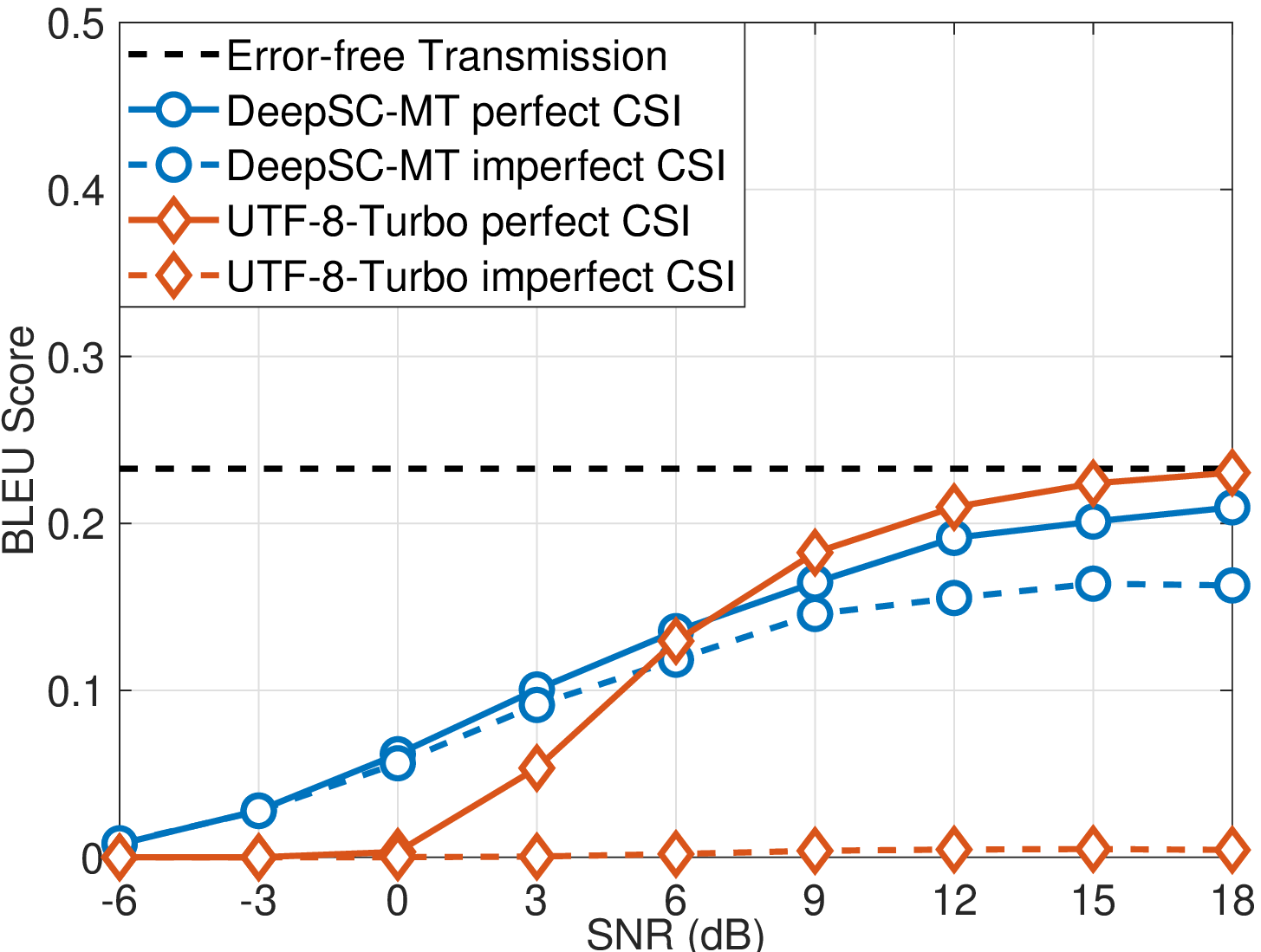}
		\label{fig:zh-en-2}
    }\hspace{-3mm}
    \subfigure[Rician Channels.]{
   		 	\includegraphics[width=60mm]{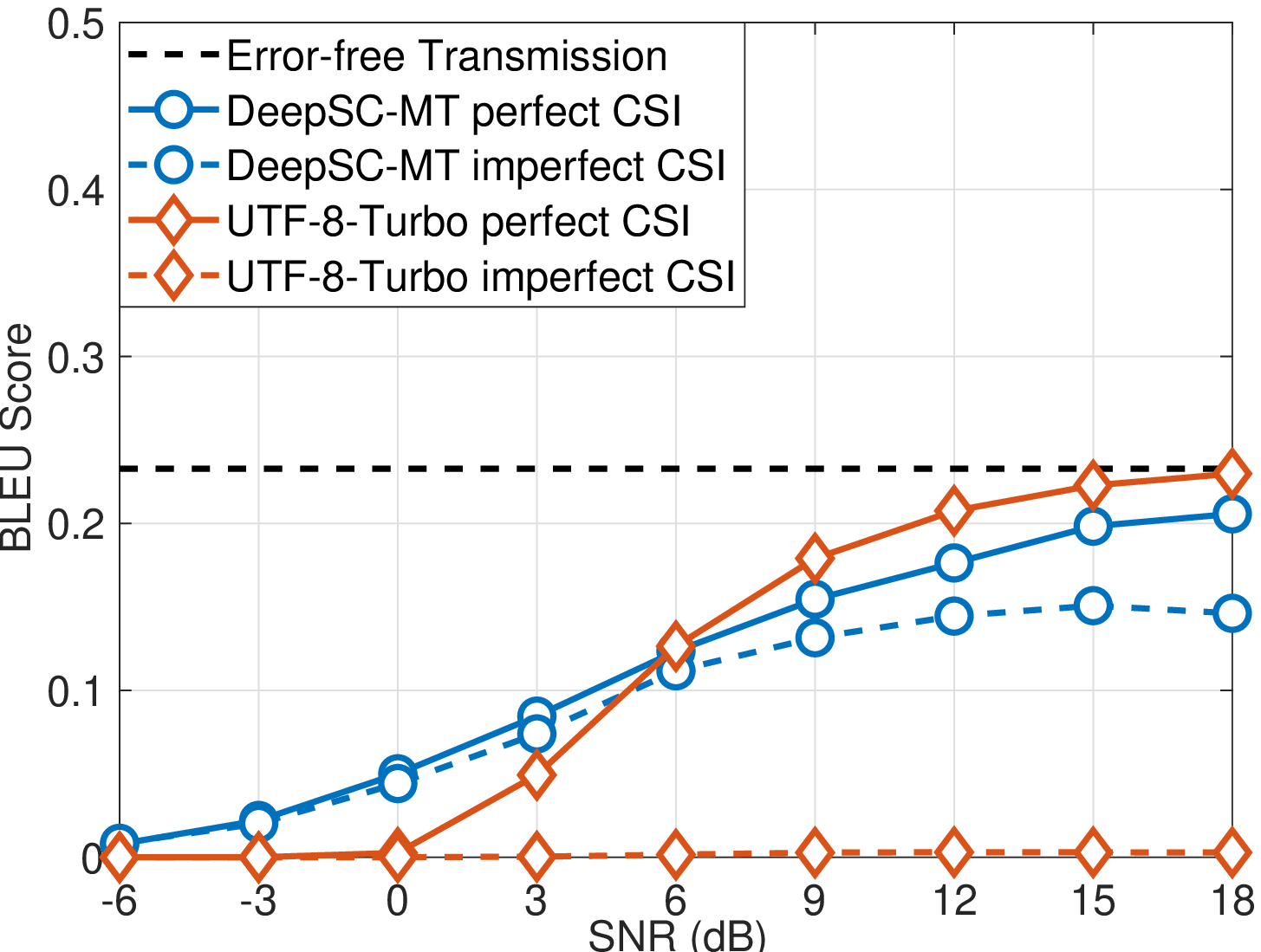}
		\label{fig:zh-en-3}
    }\hspace{-9mm}
	\caption{BLEU score comparison between DeepSC-MT and UTF-8-Turbo with BPSK for Chinese-to-English under  {AWGN channels, Rayleigh channels, and Rician channels.}}
	\label{fig:zh-en}
\end{figure*}

The BLEU score performance comparison for different channels on English-to-Chinese is reported in Fig.~\ref{fig:en-zh} and on Chinese-to-English is shown in Fig.~\ref{fig:zh-en}. From Fig.~\ref{fig:en-zh},  on English-to-Chinese over different channels,  the DeepSC-MT outperforms the UTF-8-Turbo with QPSK at the low SNR regimes over AWGN, as well as at all SNR regimes over fading channels. More inaccurate CSI decreases BLEU score for both systems, in which the DeepSC-MT outperforms the benchmark and retains its high robustness to imperfect CSI. On Chinese-to-English over fading channels in Fig.~\ref{fig:zh-en}, the DeepSC-MT performs well except at the high SNR regimes. Although the UTF-8-Turbo in BSPK has a higher BLEU score than DeepSC-MT as SNR increases, it performs worse than DeepSC-MT at all SNR regimes w.r.t. imperfect CSI. 

\begin{figure*}[!t]
	\centering
	\hspace{-9mm}
	\subfigure[AWGN Channels.]{
			\includegraphics[width=60mm]{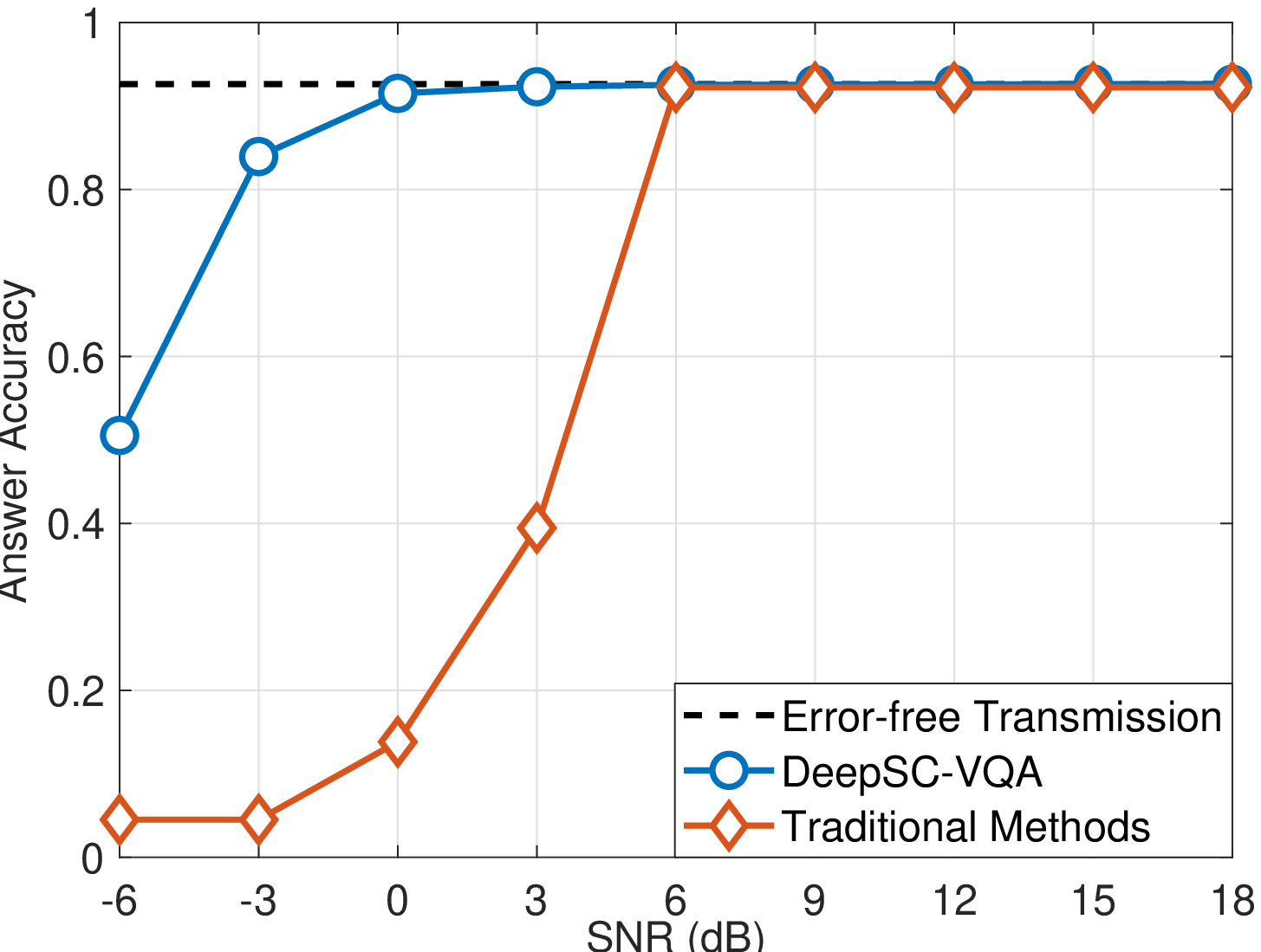}
		\label{fig:vqa-1}
	}\hspace{-3mm}
    	\subfigure[Rayleigh Channels.]{
		 	\includegraphics[width=60mm]{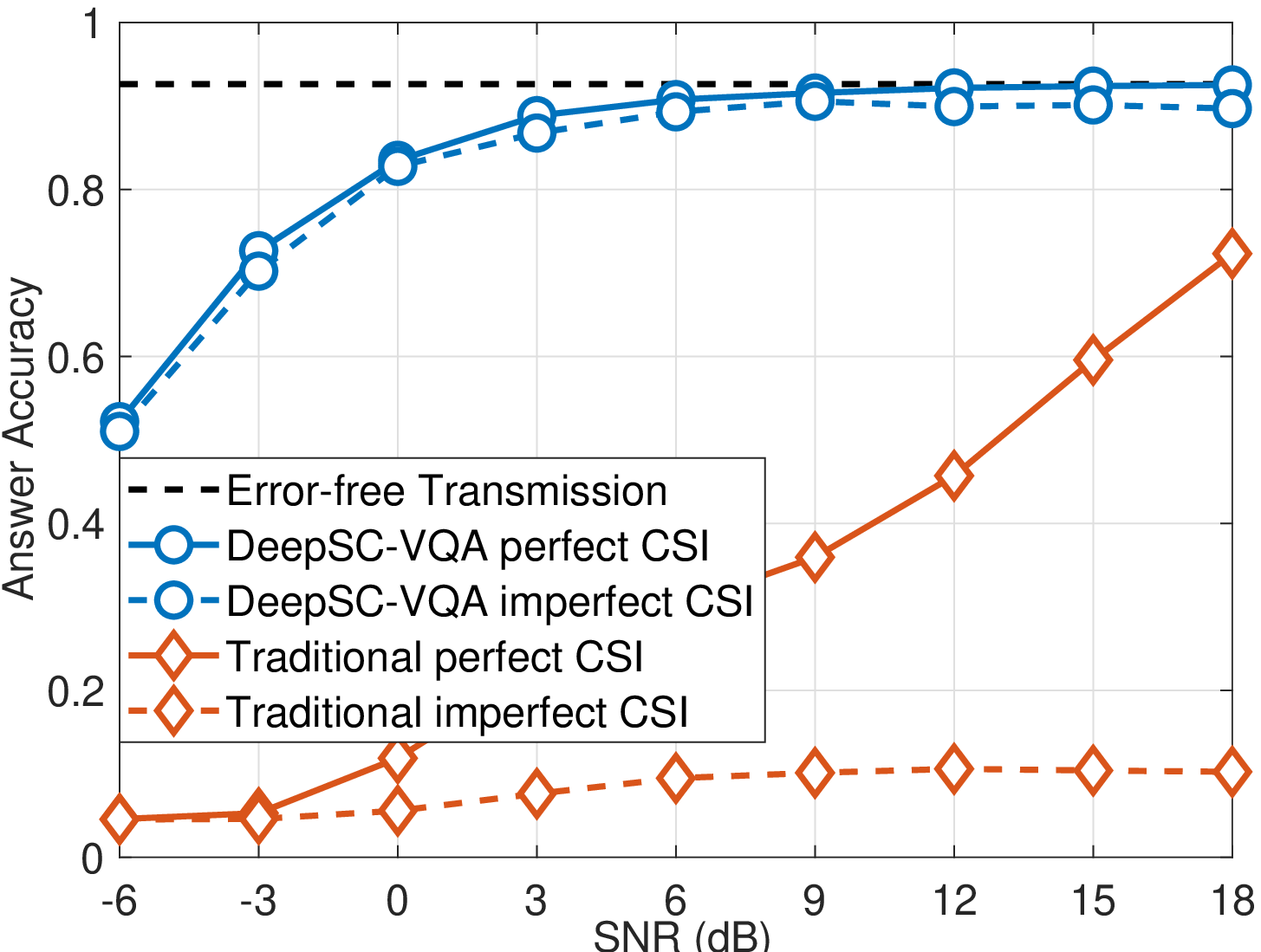}
		\label{fig:vqa-2}
    }\hspace{-3mm}
    \subfigure[Rician Channels.]{
   		 	\includegraphics[width=60mm]{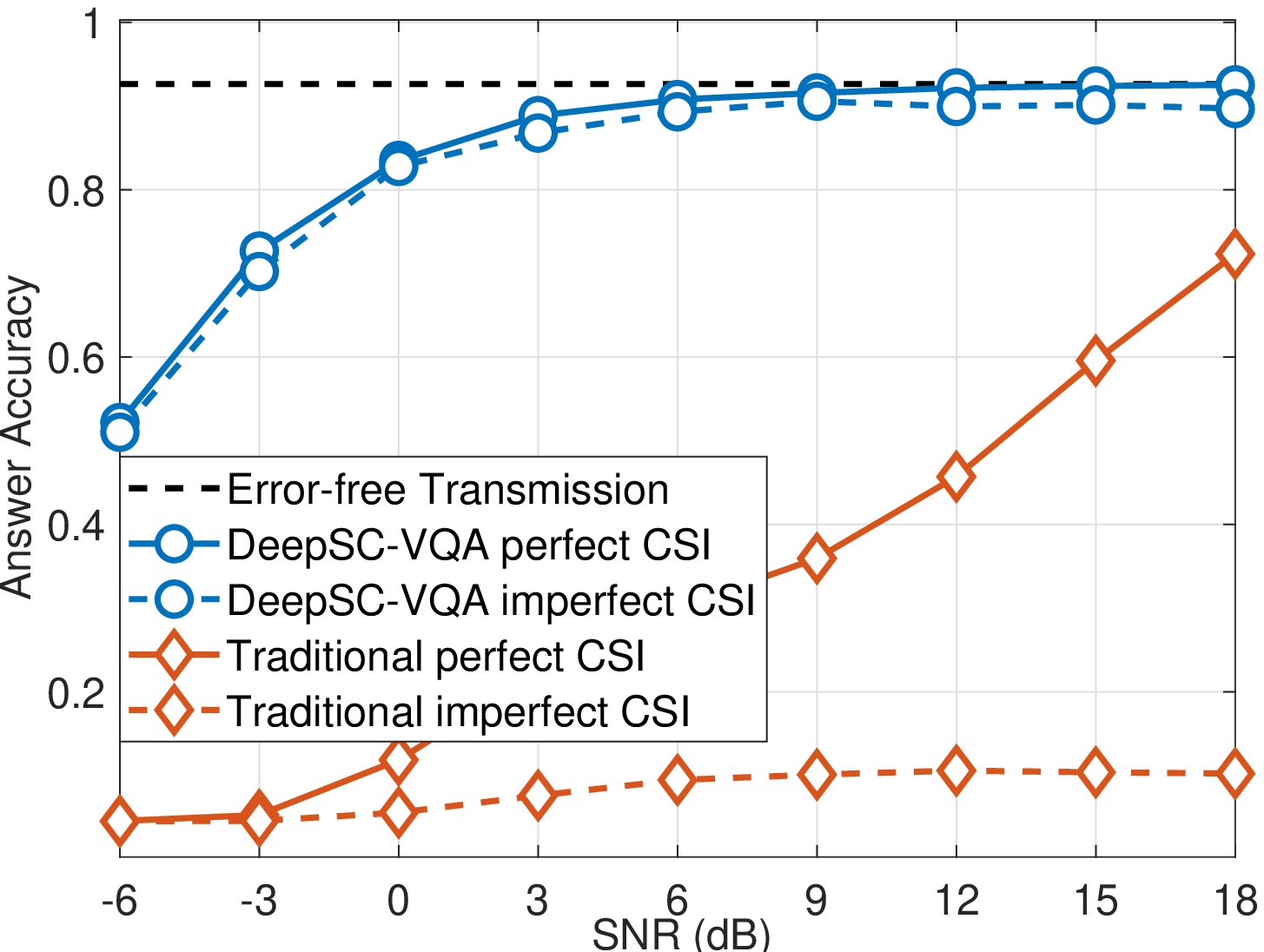}
		\label{fig:vqa-3}
    }\hspace{-9mm}
	\caption{Answer accuracy comparison between DeepSC-VQA and traditional methods, including UTF-8-Turbo with BPSK for text and JPEG-LDPC with 8-QAM for image, in which different channels are considered. }
	\label{fig:vqa}
\end{figure*}

\begin{figure*}[!t]
	\centering
	\hspace{-9mm}
	\subfigure[Image Retrieval.]{
			\includegraphics[width=60mm]{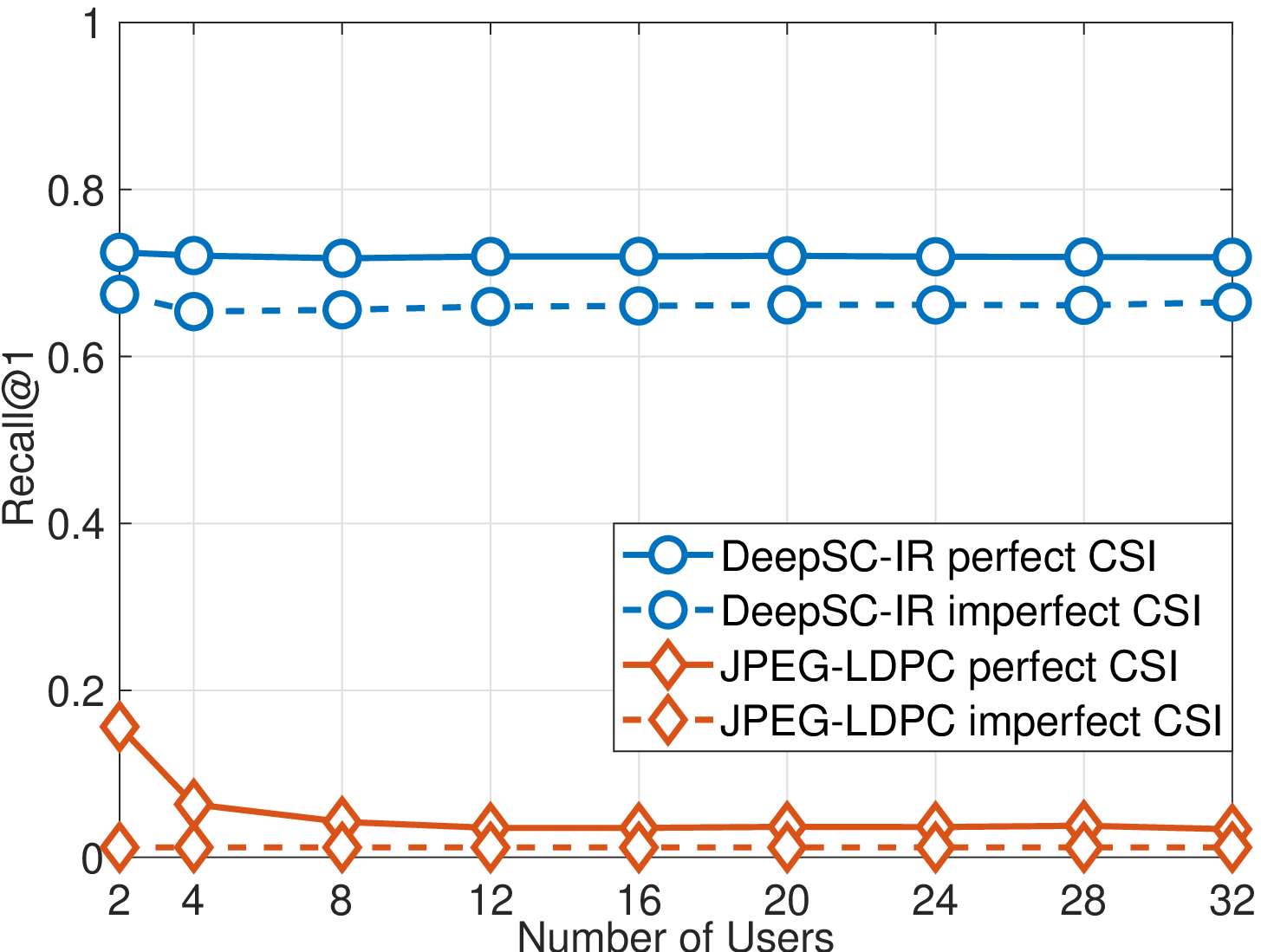}
		\label{fig:diffusers-1}
	}\hspace{-3mm}
    	\subfigure[Machine Translation.]{
		 	\includegraphics[width=60mm]{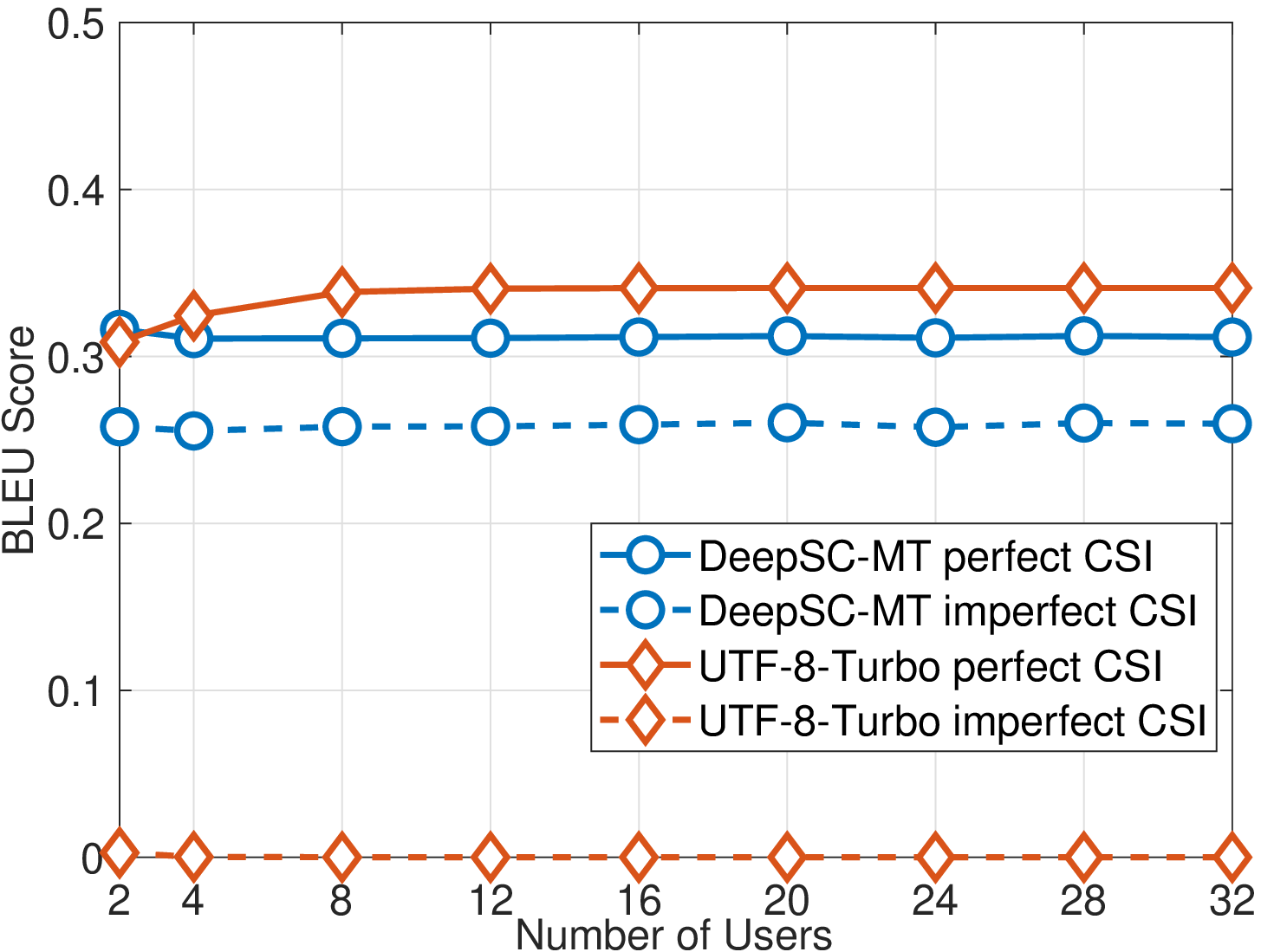}
		\label{fig:diffusers-2}
    }\hspace{-3mm}
    \subfigure[VQA.]{
   		 	\includegraphics[width=60mm]{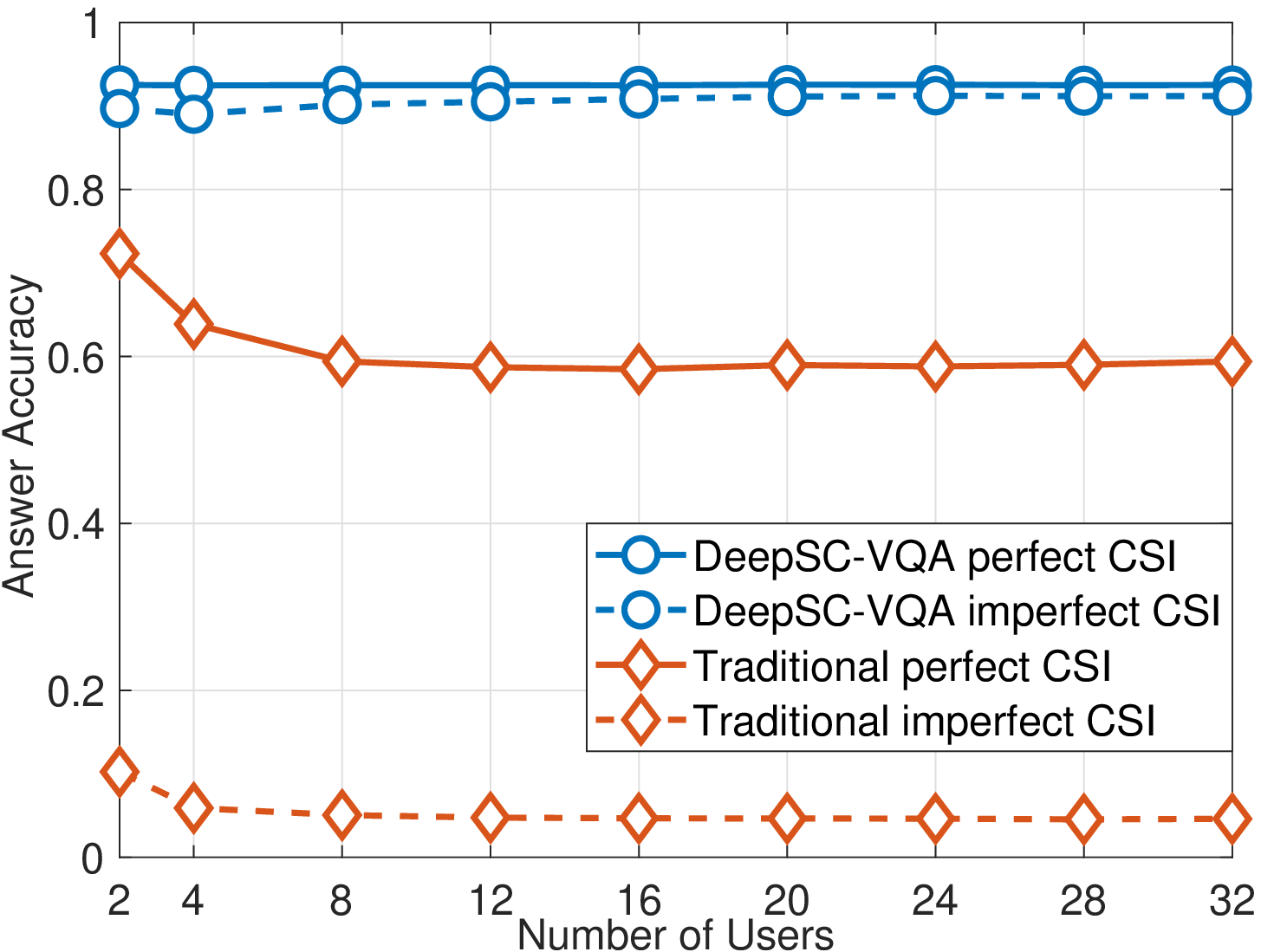}
		\label{fig:diffusers-3}
    }\hspace{-9mm}
	\caption{Recall@1, BLEU score, and answer accuracy comparisons versus the number of users over Rician channel with SNR=18dB.}
	\label{fig:diffusers}
\end{figure*}

\subsection{Multimodal Multi-User Semantic Communication}
The answer accuracy performance comparison for VQA task over different channels is presented in Fig.~\ref{fig:vqa}, in which the benchmark consists of UTF-8-Turbo with BPSK for text and JPEG-LDPC with 8-QAM for image. The DeepSC-VQA outperforms the benchmark at the low SNR regimes over the AWGN channels and at all SNR regimes over fading channels. In particular,   the DeepSC-VQA achieves the upper bound at approximate SNR=9dB over fading channels.  The answer accuracy of benchmark considerably decreases from the AWGN to fading channels for benchmarks but experiences only little performance degradation at the low SNR regimes and no performance loss at the high SNR regimes for DeepSC-VQA. Similarly, for imperfect CSI, the robustness of DeepSC-VQA is also better than that of benchmark with more than 24dB gain at 0.7 answer accuracy. This also verifies the effectiveness of the design of semantic decoder of DeepSC-VQA.

\begin{table*}[!t]
\caption{The number of transmitted symbols comparison between multi-user semantic communication systems and traditional source-channel communication systems.}
\label{tab:1}
\centering
\begin{tabular}{ c|c|c|c|c } 
\toprule
Task & Dataset & Methods & \makecell[c]{Average Number of Transmitted \\Symbols for One Image or One Word} & Ratio \\
\midrule
\multirow{4}{*}{Image Retrieval} &  Cars196 & \multirow{4}{*}{DeepSC-IR $/$ JPEG-LDPC with 8-QAM} & $128 / 499,920$  & $0.02$\%\\
&  CUB-200-2011 &   & $128 / 247,312$ & $0.05$\% \\
& In-Shop Clothes &  & $128 / 60,696$ & $0.21$\%\\
& Stanford Online Products &  & $128 / 174,808$ & $0.07$\%\\
\midrule
\multirow{2}{*}{Machine Translation} & English-to-Chinese &DeepSC-MT $/$ UTF-8-Turbo with QPSK & $77/76$ & $101.31$\%  \\
& Chinese-to-English & DeepSC-MT $/$ UTF-8-Turbo with BPSK & $77/68$ & $113.23$\% \\
\midrule
\multirow{2}{*}{VQA} & CLEVR: Text &  DeepSC-VQA $/$ UTF-8-Turbo with BPSK &$77/152$ & $50.66$\% \\
&  CLEVR: Image  & DeepSC-VQA $/$ JPEG-LDPC with 8-QAM & $25,216/55,624$ & $45.33$\% \\
\bottomrule
\end{tabular}
\end{table*}

\subsection{Different Number of Users}
In Fig.~\ref{fig:diffusers},  different tasks versus the different number of users are compared. All proposed methods perform steadily as the number of users increases but the benchmarks experience performance improvement or degradation. The difference in performance trends between benchmarks are because of the gains from channel coding and low-order modulation methods.  Both for image retrieval task and VQA task, the DeepSC-IR and DeepSC-VQA outperform their benchmarks at Recall@1 and at answer accuracy, respectively, in which the performance at Recall@1 and answer accuracy of benchmarks decrease first and achieve floor as the number of users increases. For the machine translation task, the BLEU score of the benchmark increases with the number of users, making the benchmark outperform DeepSC-MT with respect to perfect CSI. Besides, for imperfect CSI, all proposed semantic communication systems outperform the corresponding benchmarks with relatively little performance degradation.

\subsection{Number of Transmitted Symbols}
The numbers of transmission symbols for different methods are compared in Table \ref{tab:1}. For image transmission, the proposed multi-user semantic communication systems significantly decrease the number of transmission symbols, especially for the image retrieval task with the DeepSC-IR only transmitting 0.02\% symbols of the benchmarks for one image. For text transmission, although the proposed methods transmit a similar or slightly more number of symbols compared with the benchmark in machine translation task,  they achieve approximately 50\% saving in the numbers of symbols when the benchmark employs a lower order modulation in the VQA task. This suggests that the proposed multi-user semantic communications can decrease the transmission delay with a lower number of transmission symbols and hence are suitable for lower latency scenarios.

\begin{table*}[!t]
\caption{Computational complexity comparison between multi-user semantic communication systems and traditional source-channel communication systems.}
\label{tab:2}
\centering
\begin{tabular}{ c|c|c|c|c } 
\toprule
\multirow{2}{*}{Task} & \multirow{2}{*}{Dataset} & \multirow{2}{*}{Methods} & \multicolumn{2}{c}{Computational Complexity}  \\
\cmidrule{4-5}
& & & Additions  &  Multiplications\\
\midrule
\multirow{4}{*}{Image Retrieval} &  Cars196 & \multirow{4}{*}{DeepSC-IR $/$ JPEG-LDPC with 8-QAM} &  $8.2\times 10^5/9.0\times 10^9$ & $8.2\times 10^5/1.7\times 10^{10}$\\
&  CUB-200-2011 &   & $8.2\times 10^5/4.4\times 10^9$  & $8.2\times 10^5/8.4\times 10^9$\\
& In-Shop Clothes &  & $8.2\times 10^5/1.0\times 10^9$ & $8.2\times 10^5/2.1\times 10^9$\\
& Stanford Online Products &  & $8.2\times 10^5/3.1\times 10^9$ & $8.2\times 10^5/6.0\times 10^9$\\
\midrule
\multirow{2}{*}{Machine Translation} & English-to-Chinese &DeepSC-MT $/$ UTF-8-Turbo with QPSK & $5.9\times 10^5/1.0\times10^5$ & $5.9\times 10^5/1.6\times 10^5$  \\
& Chinese-to-English & DeepSC-MT $/$ UTF-8-Turbo with BPSK & $5.9\times 10^5/4.5\times 10^4$ & $5.9\times 10^5/7.3\times 10^4$ \\
\midrule
\multirow{2}{*}{VQA} & CLEVR: Text &  DeepSC-VQA $/$ UTF-8-Turbo with BPSK & $5.9\times 10^5/1.0\times10^5$ & $5.9\times 10^5/1.6\times 10^5$ \\
&  CLEVR: Image  & DeepSC-VQA $/$ JPEG-LDPC with 8-QAM & $1.6\times 10^8/1.0\times 10^9$ & $1.6\times 10^8/1.9\times 10^9$ \\
\bottomrule
\end{tabular}
\end{table*}

\subsection{Computational Complexity}
The computational complexity for different methods\footnote{We only analyze the complexity of channel coding for both methods because the other parts are shared in both methods and the complexity of source coding is low and can be omitted.} is compared in Table \ref{tab:2}. For image transmission, all of the proposed methods have a lower computational complexity than traditional methods, in which the complexity of DeepSC-IR can decrease by more than one order of magnitude. For text transmission, the proposed DeepSC-MT shows a similar computational complexity in English transmission but has a slightly higher computational complexity in the Chinese transmission compared to the benchmarks. Such a slightly higher computational complexity can provide robustness to noise in low SNRs. This suggests that the proposed multi-user semantic communication systems achieve lower power consumption when transmitting a large size of data.

\begin{figure*}
    \centering
    \includegraphics[width=150mm]{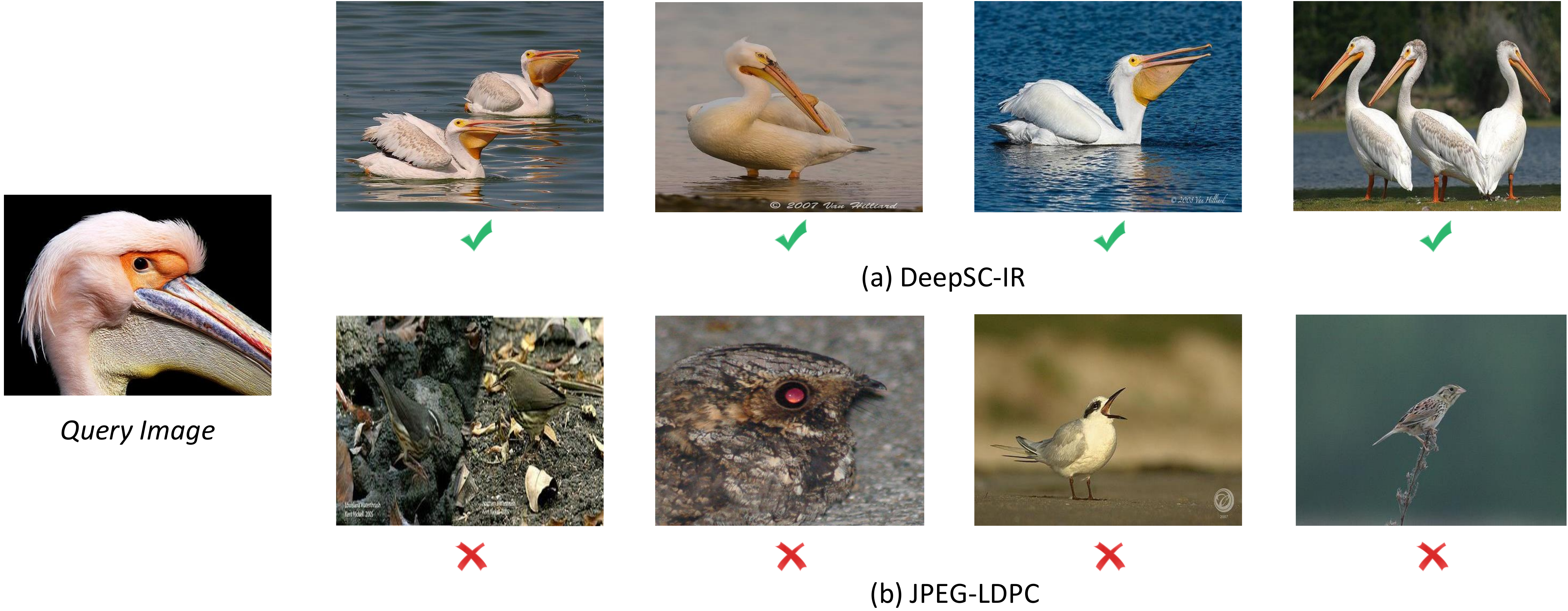}
    \caption{Part of the results for image retrieval at 18 dB over Rician Channels.}
\end{figure*}

\begin{table*}[!t]
\caption{Part of the results for machine translation on Chinese to English at 18 dB over Rician Channels.}
\centering
\begin{tabular}{ |c|c| } 
\toprule
Transmission Sentence & \begin{CJK}{UTF8}{gbsn} 
与此同时,过去二十年来卡塔尔想尽办法扩张其影响力并已经收到成效,该国的吸引力已经蔚为可观。\end{CJK}\\
\midrule
Reference Translated Sentence & \makecell[c]{Meanwhile, Qatar's diligent efforts to expand its influence over the last two decades have paid off, \\ with the country developing considerable power of attraction.} \\
\midrule
DeepSC-MT perfect CSI & \makecell[c]{Meanwhile, over the past two decades, Qatar has been able to expand its influence \\ by doing everything it can,  and has been successful, and the country's appeal is already strong.}\\
\midrule
DeepSC-MT imperfect CSI & \makecell[c]{Meanwhile, over the past two decades, Qatar has been able to reap the benefits of its efforts \\ to expand its reach, and the country's appeal is already strong.}\\
\midrule
UTF-8-Turbo with BPSK perfect CSI & \makecell[c]{At the same time, over the past two decades Qatar had tried to expand its influence \\ and had already borne fruit, and the country's appeal had become significant.}\\
\midrule
UTF-8-Turbo with BPSK imperfect CSI & \makecell[c]{Could not decode}\\
\bottomrule
\end{tabular}
\end{table*}

\begin{figure*}
    \centering
    \includegraphics[width=150mm]{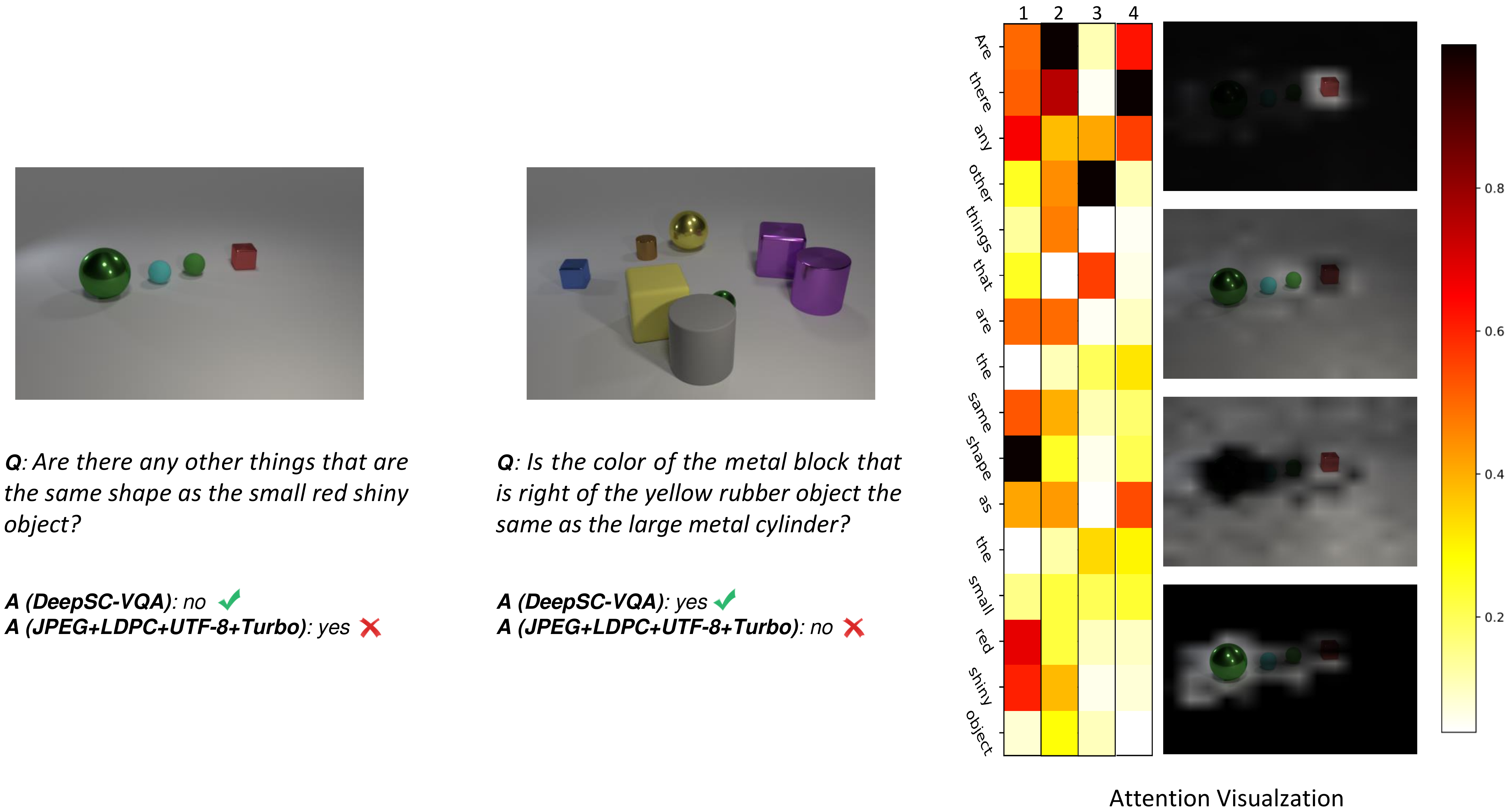}
    \caption{Part of results for the VQA task and visualized attention for layer-wise Transformer at 18 dB over Rician Channels.}
    \label{deepsc-vqa-visual-results}
\end{figure*}

\subsection{Visualization Results}
The visualized results for the considered tasks including image retrieval, machine translation, and VQA are shown in Fig. 10, Table III, and Fig. 11, respectively. Fig. 10 shows \textit{top-4} similar image retrieval results for DeepSC-IR and JPEG-LDPC at 18 dB over Rician channels. The proposed DeepSC-IR returns similar images successfully with the query image but the traditional method fails due to the destroyed received image. Table III provides the received translation results on Chinese-to-English. The proposed DeepSC-MT demonstrates reasonable translations results in both scenarios with perfect and imperfect CSI, but the traditional method fails to convey the sentence when CSI cannot be estimated exactly.

Fig. \ref{deepsc-vqa-visual-results} shows the results of the VQA task and the attention visualizations for the layer-wise Transformer. The proposed DeepSC-VQA correctly answers the question. In the attention visualizations, the proposed DeepSC-VQA can effectively query the key regions in the image layer by layer with the received semantic image and text information. Specifically, the words ``\textit{shape, red tiny}'' in the sentence has a higher magnitude in the first layer, finding the key red tiny object in the image. The second layer highlights the words ``\textit{are there other}'' to find other objects in the image and neglect the red tiny object. The third and fourth layers double-check the other objects with the red tiny object to give the final answer. 

\section{Conclusions}
In this paper, we have explored task-oriented multi-user semantic communications to transmit data with single-modality and multiple modalities, respectively. We considered two single-modal tasks, image retrieval and machine translation, as well as one multimodal task, visual question answering (VQA). In this context, we have proposed three Transformer based transceivers, DeepSC-IR, DeepSC-MT, and DeepSC-VQA, which share the same transmitter structures but with different receiver structures. Each transceiver is trained jointly by the proposed training algorithm. In addition, all of the proposed multi-user semantic communication systems were found to outperform the traditional ones in the low SNR regimes and provide graceful performance degradation with imperfect CSI. For both image retrieval and VQA tasks, the proposed DeepSC-IR and DeepSC-VQA can provide more than 18 dB gain and reduce by more than 50\% the number of transmission symbols and computational complexity compared to traditional communications. In particular, compared with traditional methods, DeepSC-IR only needs 1\textperthousand \, transmission symbols on average and decreases the complexity by more than one order of magnitude. As we result, we can conclude that multi-user semantic communication systems are an attractive alternative to traditional communication systems for particular tasks.




%





\ifCLASSOPTIONcaptionsoff
  \newpage
\fi



\bibliographystyle{IEEEtran}
\bibliography{IEEEabrv, reference.bib}

\end{document}